\documentclass[twocolumn]{aastex62}
\pdfoutput=1
\usepackage{hyperref}

\renewcommand{\baselinestretch}{1.1}

\usepackage{natbib}
\usepackage{gensymb}
\usepackage{epstopdf}
\usepackage{cleveref}
\usepackage[multiple]{footmisc}
\crefname{paragraph}{\S}{\S\S}
\shorttitle{Stanek Field Study} 
\shortauthors{S. K. Terry, et al.}

\begin{document}
\small
\title{Comparing Observed Stellar Kinematics and Surface Densities in a Low Latitude Bulge Field to Galactic Population Synthesis Models}

\author{Sean K. Terry}
\affiliation{Department of Physics, Catholic University of America, 620 Michigan Ave., N.E. Washington, DC 20064, USA}
\affiliation{Code 667, NASA Goddard Space Flight Center, Greenbelt, MD 20771, USA}

\author {Richard K. Barry}
\affiliation{Code 667, NASA Goddard Space Flight Center, Greenbelt, MD 20771, USA}

\author{David P. Bennett}
\affiliation{Code 667, NASA Goddard Space Flight Center, Greenbelt, MD 20771, USA}
\affiliation{Department of Astronomy, University of Maryland, College Park, MD 20742, USA}

\author {Aparna Bhattacharya}
\affiliation{Code 667, NASA Goddard Space Flight Center, Greenbelt, MD 20771, USA}
\affiliation{Department of Astronomy, University of Maryland, College Park, MD 20742, USA}

\author{Jay Anderson}
\affiliation{Space Telescope Science Institute, 3700 San Martin Drive, Baltimore, MD 21218, USA}

\author{Matthew T. Penny}
\affiliation{Department of Physics and Astronomy, Louisiana State University, Baton Rouge, LA 70803 USA}

\correspondingauthor{S. K. Terry}
\email{41terry@cua.edu}

\begin{abstract}

\small We present an analysis of Galactic bulge stars from \textit{Hubble Space Telescope} (HST) \textit{Wide Field Camera 3} (WFC3)  observations of the Stanek window (l,b=[0.25,-2.15]) from two epochs approximately two years apart. This dataset is adjacent to the provisional \textit{Wide-field Infrared Survey Telescope} (WFIRST) microlensing field. Proper motions are measured for approximately 115,000 stars down to 28th mag in V band and 25th mag in I band, with accuracies of 0.5 mas yr$^{-1}$ (20 km s$^{-1}$)  at I $\approx$ 21. A cut on the longitudinal proper motion $\mu_l$ allows us to separate disk and bulge populations and produce bulge-only star counts that are corrected for photometric completeness and efficiency of the proper-motion cut. The kinematic dispersions and surface density in the field are compared to the nearby SWEEPS sight-line, finding a marginally larger than expected gradient in stellar density. The observed bulge star counts and kinematics are further compared to the Besan\c{c}on, Galaxia, and GalMod Galactic population synthesis models. We find that most of the models underpredict low-mass bulge stars by $\sim$33\% below the main-sequence turnoff, and upwards of $\sim$70\% at redder J and H wavebands. While considering inaccuracies in the Galactic models, we give implications for the exoplanet yield from the WFIRST microlensing mission.\bigskip

 \vspace{10mm}
\end{abstract}

\section{Introduction} \label{para:1}
\begin{figure}[ht!]\label{fig:WFIRST_Fields}
\figurenum{1}
\epsscale{1.2}
\plotone{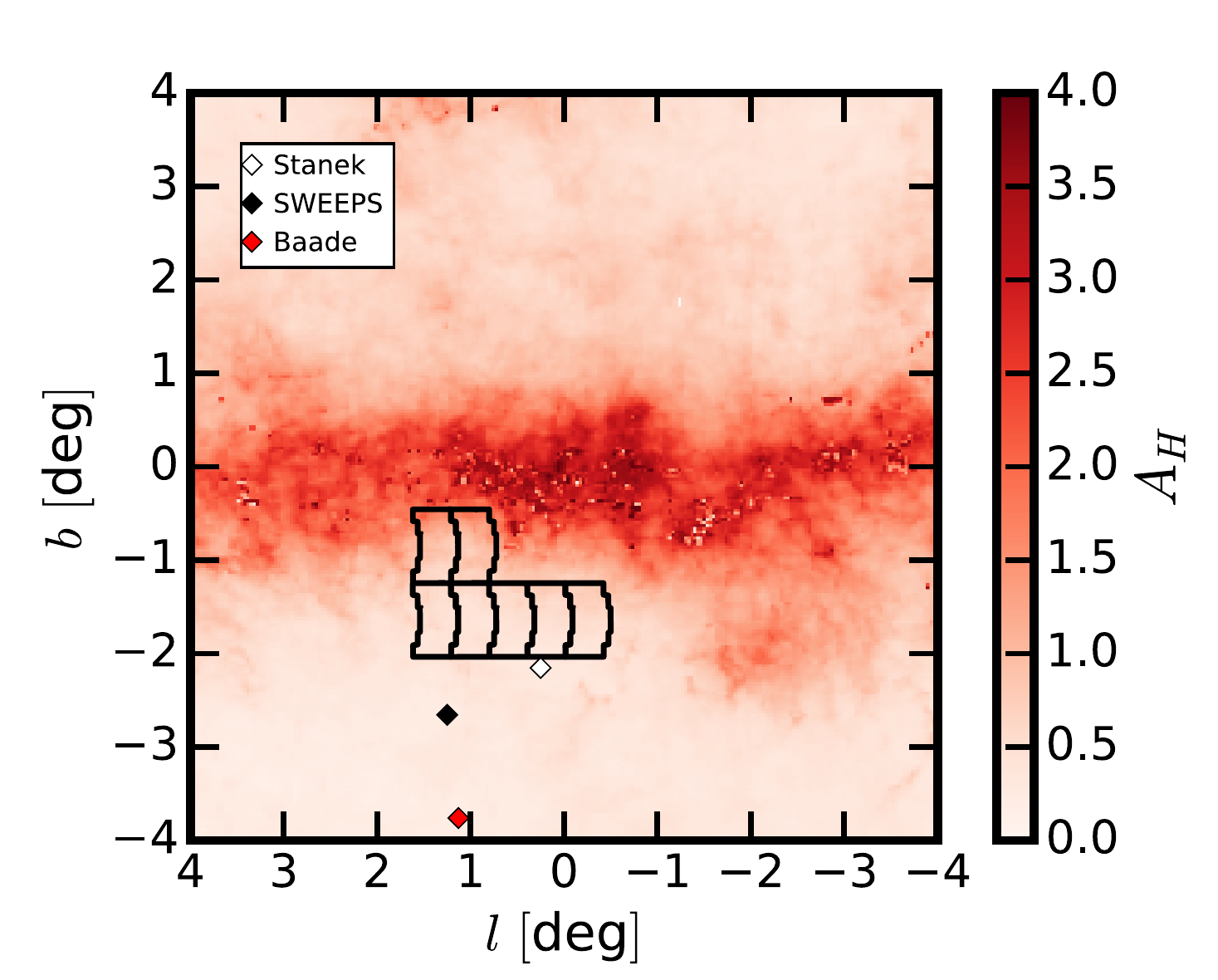}
\caption{\scriptsize{Current Cycle 7 WFIRST microlensing survey footprint \citep{penny:2019aa}, with H-band extinction map from \cite{gonzalez:2012aa} overplotted. The field locations for Stanek, SWEEPS, and Baade have been enlarged for visibility.}}
\end{figure}

\indent The stellar environment toward the center of the Milky Way is of fundamental importance in our understanding of Galactic evolution, structure, and dynamics.  Prior studies have focused on the important task of measuring the red clump (RC) giant stars toward the center of the galaxy \citep{stanek:1994aa, stanek:1997aa, nataf:2010aa} as these stars are very good standard candles to probe the Galactic bulge population. The central bulge in our Galaxy is kinematically and chemically distinct from other structures and is of great current interest as new observational techniques and technologies enable deeper and more complete assessment of its components. Until recently, the bulge was thought to be nearly homogeneously old and metal rich (\cite{ortolani:1995aa} and references therein). Resolution and characterization of individual stars in the bulge into luminosity functions (LF) have shown this not to be the case (c.f. \cite{Calamida:2015aa} and references therein).\\
\indent Surface densities are important in the study of stellar populations to disentangle different kinematic groups and to identify coeval subgroups.  To be of value as an LF, stars in the sample must be identified as unambiguously point-like with unblended color.  A precise, deep LF for stars in the Galactic bulge is difficult to produce, especially at low latitudes, due to several factors including crowding, extinction and contamination of the sample with disk stars.\\
\indent Gravitational microlensing requires the exceedingly precise alignment of a foreground mass (the lens) with a background star (the source) (cf. \cite{Gaudi:2012aa} for a  review). A productive microlensing survey thus requires very dense stellar fields to supply a suitable microlensing event rate and optical depth \citep{Mao:2008aa}. Measurement of microlensing parallax, lens star color and/or detection of finite source effects from a resolved source star are needed to yield the masses - critical for tests of planet formation theories \citep{suzuki:2016aa}.  A thorough understanding of the underlying source star distribution in the form of a deep LF is of principal importance in the prediction of the microlensing event rate and the exoplanet detection efficiency.\\
\indent This detailed knowledge will be needed prior to the launch of WFIRST in order to optimize the microlensing mission's scientific yield \citep{Spergel:2015aa}, and, indeed, forms the basis of the mission's exoplanet success criterion \citep{Yee:2014aa}. Currently predicted detection rates have large uncertainties for several apparent reasons.  Firstly, the LFs that exist at present have mostly been conducted in visible light. The WFIRST mission will observe the bulge in the near infrared (near-IR) J and H bands mostly, and the observational fields under consideration are heavily extincted in visible light.  Consequently, stellar densities, microlensing event rates and planet detection efficiencies are all extrapolated from visible observations. The lack of deep near-IR bulge stellar classification studies thus contributes to the uncertainty in expected yields for the mission. Secondly, while deep bulge luminosity functions do exist, they are focused on sight-lines that are distant from projected WFIRST fields and older studies are contaminated by foreground stars (\cite{Holtzman:1998aa}, \cite{zoccali:2000aa}).  Importantly, Stanek's field, the sight-line that is closest to the proposed WFIRST exoplanet survey field has been observed using HST at several epochs, and can now be analyzed in the context of the microlensing campaign itself.\\
\indent Figure \ref{fig:WFIRST_Fields} shows the most up-to-date planned WFIRST microlensing survey fields from \cite{penny:2019aa} (hereafter P19) overlayed on an H-band extinction map near the Galactic center \citep{gonzalez:2012aa}. The white marker indicates the area of stars analyzed in this work and the approximate center of the Stanek field. The black and red markers indicate the approximate locations of the SWEEPS field and Baade's window respectively.\\
\indent Several studies have incorporated these HST Cycle 17 images in recent years. Age and metallicity estimates of globular clusters observed in the program were performed by \cite{Milone:2012aa}, \cite{Lagioia:2014aa}, \cite{Calamida:2014ab}, and \cite{Baldwin:2016aa}. A broad study of the star formation rate (SFR) and initial mass function (IMF) was also conducted by the GO PI's \citep{Gennaro:2015aa} using predominantly the OGLE29 field images. The Star Formation History derived in the Stanek and SWEEPS fields using the WFC3 photometry was conducted by \cite{bernard:2018aa} and found to be quite similar between the two regions. Additionally, a detailed metallicity study of the fields were made by \cite{renzini:2018aa}. They find that the most metal-poor and metal-rich components are essentially coeval and only a small fraction ($\sim$3\%) of metal-rich bulge stars are 5 Gyr or younger. There has yet to be a substantial probe of the bulge stellar kinematics and surface densities (which are most important for microlensing) in the Stanek field with regard to the WFIRST.\\
\indent The mission, launching in the mid-2020's, will discover thousands of exoplanets by microlensing (P19) and transits \citep{montet:2017aa} in this area of the Galactic bulge. The mission is also expected to discover a small number of Mars-mass free-floating planets (FFPs)\citep{sumi:2011aa, barclay:2017aa, mroz:2019aa}. A strong understanding of the population of disk (foreground and background) and bulge stars in this window is necessary.

\begin{deluxetable}{ccCrlcc}
\deluxetablecaption{Bulge field properties\label{table:1}}
\tablecolumns{6}
\tablenum{1}
\tablewidth{\columnwidth}
\tablehead{
\colhead{Field} &
\colhead{$l$} & \colhead{$b$} & \colhead{$A_I$} & \colhead{$R_{GC}$}\\
\colhead{} & \colhead{[deg]} & 
\colhead{[deg]} & \colhead{[mag]} & \colhead{[kpc]}
}
\startdata
Stanek & 0.25 & -2.15 & 1.284 & 0.32\\
SWEEPS & 1.25 & -2.65 & 1.004 & 0.43\\
Baade & 1.06 & -3.81 & 0.743 & 0.58\\
\enddata
\end{deluxetable}

\indent This paper is organized as follows. We discuss the data, its reduction and photometry in Section \ref{sec:2}. in Section \ref{para:3} we present our analysis of the astrometry and subsequent PM measurements, along with an analysis of the PM dispersion in the field and an estimation of the efficiency of the cut made on the the Galactic longitude $\mu_l$ direction. We include a brief section (\ref{sec:RotCurve}) describing the Galactic rotation curve from the measured mean PM values along the sight-line. In Section \ref{para:4} we present corrected star counts in the field, and compare these results with several empirical Galactic models. Section \ref{para:5} includes the near-IR analysis of the bulge stars, again with star count results and model comparisons.  We finish the paper in Section \ref{sec:6} with a discussion of the results and implications for the WFIRST microlensing survey.

\section{Observations and Data Reduction} \label{sec:2}
Detailed descriptions of the observations are discussed in \cite{brown:2009aa,brown:2010aa}, so only a brief overview is given here. The observations were conducted as part of the WFC3 Galactic Bulge Treasury Program\footnote[6]{\href{https://archive.stsci.edu/prepds/wfc3bulge/}{\scriptsize https://archive.stsci.edu/prepds/wfc3bulge/}} in June 2010 (GO-11664) and June 2012 (GO-12666) in various passbands; F390W, F555W, F814W in the WFC3/UVIS channel and F110W, F160W in the WFC3/IR channel. One of the main goals of the program was to gather deep images of four low-extinction fields toward the center of the Galaxy: OGLE29, Baade's Window, SWEEPS, and the Stanek Field. The program has recently published and subsequently updated (2018-06-05) their version 2 high-level science products (HLSP)\footnote[7]{\href{http://archive.stsci.edu/doi/resolve/resolve.html?doi=10.17909/T90K6R}{\scriptsize doi:10.17909/T90K6R}}. The V and I analysis in this paper began before the version 2 science products were released, therefore the reduction, photometry, and astrometry presented stands on its own here. Both reductions use the ``Kitchen Sink" (KS2) software package by Jay Anderson, as described later in this section. The outputs from both reductions show very similar results with regard to photometric and astrometric accuracy, as well as completeness measurements from artificial star tests (AST).\\
\indent The planned WFIRST microlensing fields are centered at [$l, b] \approx (0.5\degree, -1.7\degree$) with a total coverage area of 1.97deg$^2$ \footnote[8]{\href{https://wfirst.ipac.caltech.edu/sims/Param_db.html}{\scriptsize https://wfirst.ipac.caltech.edu/sims/Param\textunderscore db.html}}. The OGLE29, Baade's Window, and SWEEPS fields are all outside of this area, with the Stanek Field ($\sim2.7'$ x $2.7'$) covering a portion near the edge of the planned footprint at [$l, b] = (0.25\degree, -2.15\degree$). The OGLE29 field [$l, b] = (-6.75\degree, -4.72\degree$) lies substantially further away from the other three bulge fields, so it has been omitted from any comparisons in this paper. Table \ref{table:1} shows a comparison of the extinction in these fields and their respective projected distances from the Galactic center \citep{reid:2009aa}. The first set of F555W and F814W observations were taken on June 27, 2010 with seven total exposures in each passband for a total exposure time of 2283s (F555W) and 2143s (F814W). Four of the seven exposures in each filter were sub-pixel dithered to allow for high accuracy astrometry, and therefore only the dithered frames were used in the subsequent analysis. The second epoch of observations were taken on June 27, 2012 with the F814W filter. Total exposure time for the 2012 images was 1983s, and again only the four sub-pixel dithered images were analyzed.

\begin{figure*}\label{fig:StanekFieldNoClean}
\figurenum{2}
\plotone{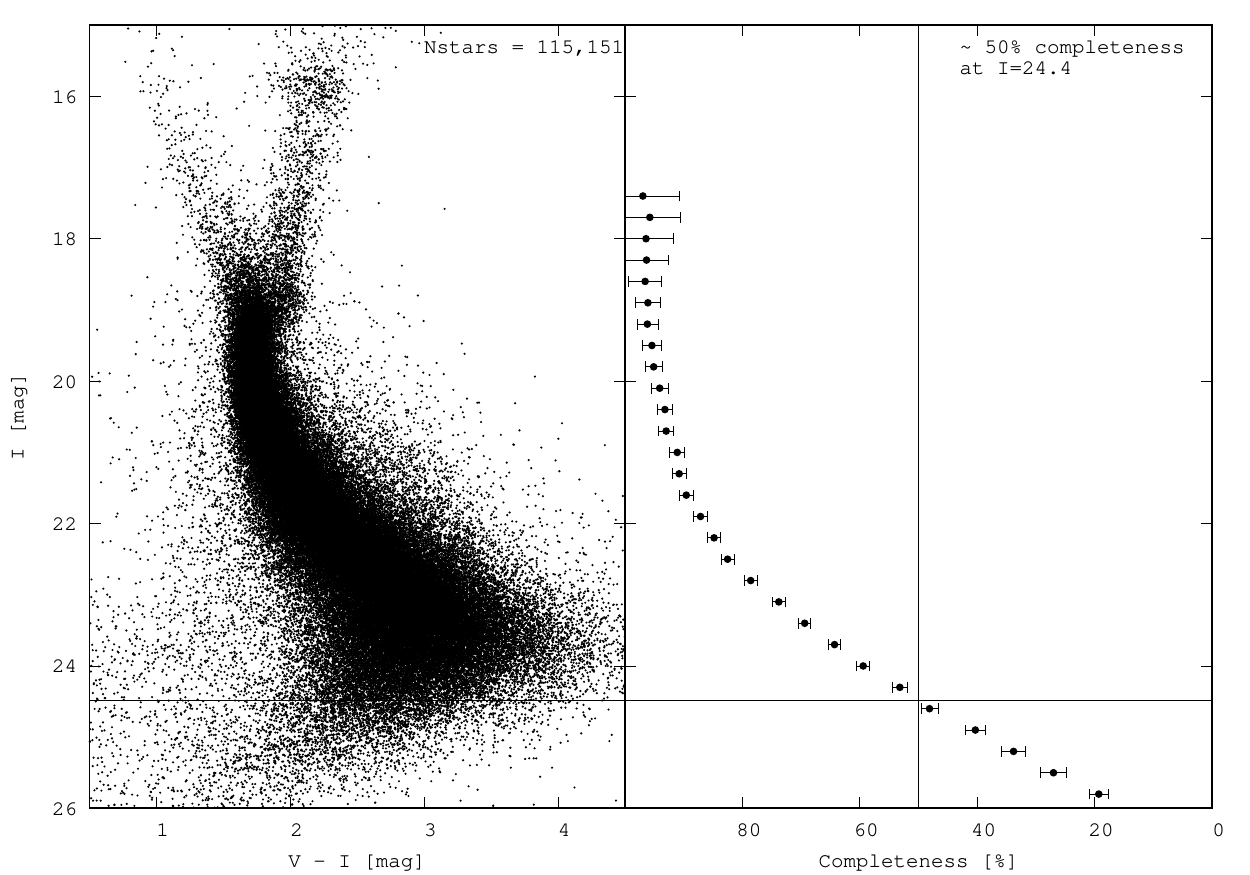}
\caption{\scriptsize{\textit{Left}: CMD of Stanek's Field stars from the 2010-2012 datasets. \textit{Right}: Completeness as a function of I mag computed from artificial star tests.}}
\end{figure*}

The data reduction was performed using a combination of \texttt{img2xym} and \texttt{KS2} \citep{Anderson:2006aa, anderson:2008aa}; for the former, a standard 9x9 PSF grid was used for each filter that accounts for spatial variation across the WFC3/UVIS detector. The routine interpolates the four closest standard PSF's near each star and uses this interpolated PSF for the individual local measurement. A perturbation to the standard PSF's was also included to account for variations in time and location of measurements. This method results in a much higher accuracy for position measurements than a library PSF from the images themselves. A distortion solution is then used to transform the position and flux measurements from the raw frame to a distortion-free sky frame. The distortion accuracy has been measured to be $\sim$1\% of a pixel \citep{Anderson:2006aa}. In this manner, the positions and fluxes are measured to very high accuracy for both epochs. Additionally, we chose to cull the dataset by the quality-of-fit parameter \textit{q}, which is a measure of the difference in the PSF-fitted flux and the aperture flux in a 5x5 pixel radius centered on each source. We used a conservative rejection threshold of \textit{q} $\geq$ 0.25 because of the moderate crowding in these frames. This results in a majority of the remaining stars having photometric rms values $\leq$ 0.1 mag, and astrometric rms values of $\leq$ 1.0 mas in both V and I band.\\
\indent The single pass nature of the routine assumes that all stars are relatively isolated, which is not the case in this moderately crowded field. This places a limit on the fullness of the overall dataset. Subsequently, the \texttt{KS2} routine was performed on the images. By running a multi-pass reduction routine, we can probe somewhat deeper and acquire well-measured flux and position values for the fainter stars. \texttt{KS2} takes as an input the catalog of well-measured bright stars produced from the initial single-pass reduction \texttt{img2xym}. The resulting photometry dataset has a contribution of well-measured bright stars (F814W $ \leq 17$) from the single-pass routine and well-measured multi-pass routine stars (F814W $ > 17$). The instrumental photometry was then transformed to the VEGAMAG system using PySynphot \citep{astropy:2013aa, astropy:2018aa}. There are chip-based variations in the zeropoint values for the WFC3/UVIS detector, which were accounted for during calibration. For the remainder of this text, F814W I band (simply `I') and F814W V band (simply `V') are reported throughout.\\
\indent We combined the photometry and astrometry of the 2010 and 2012 datasets using a method analogous to equation's \ref{eq:1} and \ref{eq:2} to get a final dataset of 115,151 stars. The left panel of Figure \ref{fig:StanekFieldNoClean} shows the I, (V - I) CMD of Stanek field stars, while the right panel shows the photometric completeness from artificial star tests conducted in section \ref{sec:ArtStar}. A clear population of foreground disk stars is seen as an un-evolved, blue branch at bright magnitudes. The stars in this branch as well as the older, red giant branch (RGB) above the main sequence turnoff (MSTO) are used as tracer objects to map the PMs. A clear population of Red-Clump Giants (RC) is also apparent at magnitude $I\approx 15.8$. The redder `shoulder' following the main sequence (MS) at $2.0 < V - I < 4.0$ and I mag $>$ 19 is comprised primarily of foreground (and background) disk stars. These disk stars overlap the bulge population substantially in color-magnitude space and become increasingly ambiguous as the distances to disk stars approach the `beginning' of the bulge population along the sight-line. Further descriptions of this and the methodology for generating a `clean' bulge sample are given in section \ref{para:3} and subsections within.

\subsection{\normalfont{Artificial Star Tests}} \label{sec:ArtStar}
Several artificial star tests (AST) were performed using approximately 200,000 fake stars in order to characterize the photometric completeness, PM accuracy, as well as errors in the reduction routine. The tests were conducted using an artificial star mode within the routine \texttt{KS2} to accept an input list of artificial star positions, magnitudes and colors. The colors and magnitudes of fake stars were estimated by calculating the loci of each point along the MS of the real CMD including Gaussian noise around each source. Artificial stars were added to each image one by one and in a `tile by tile' pattern, adding and measuring several synthetic stars in each tile of size $\sim$120 x 120 pixels. This method is useful for avoiding major effects from crowding, which can be substantial in deep bulge field images.\\
\indent The output art-star positions and fluxes were then compared with the input art-star files. A star was considered found if it passed the following criteria:\\
\begin{equation}\label{eq:1}\sqrt{(X_{\textrm{out}} - X_{\textrm{in}})^2 + (Y_{\textrm{out}} - Y_{\textrm{in}})^2} \leq 0.50\textrm{pix}, \end{equation}\vspace{0.5mm}
\begin{equation}\label{eq:2}\left| \textrm{mag}_{\textrm{in}} - \textrm{mag}_{\textrm{out}} \right| \leq 0.50\textrm{mag},
\end{equation}

where `in' and `out' denote input star and output star respectively. Additional art-star tests were performed to estimate the PM accuracy by adopting the method of \cite{Calamida:2015aa}. Similar results were found, with somewhat lower accuracy due to the limited number of sub-pixel dithered images relative to SWEEPS. The dispersion of recovered PMs increases with dimmer magnitudes as expected, and with a measured accuracy better than 0.5 mas yr$^{-1}$ (20 km s$^{-1}$ when converting to transverse velocities at bulge distances) at magnitudes brighter than F814W $\approx$ 21 for most stars, with an accuracy $\sim$1.5 mas yr$^{-1}$ (60 km s$^{-1}$) near the 50\% photometric completeness in both V and I bands. This allows for an accurate determination of foreground disk or bulge population stars when culling the dataset based on $\mu_l$, down to faint magnitudes but clearly not as faint as SWEEPS, which is one of the deepest datasets to-date.

\begin{figure*}
\figurenum{3}
\plotone{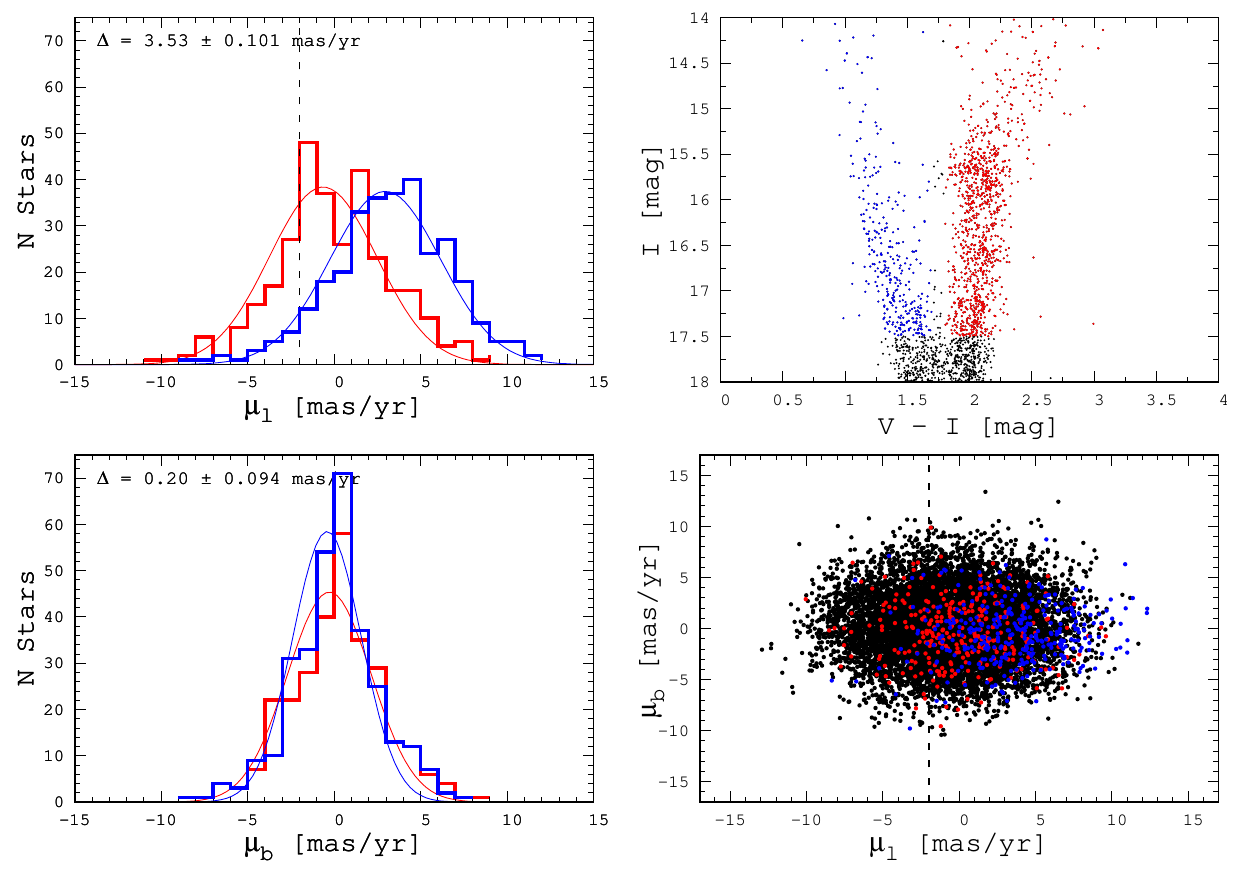}
\caption{\scriptsize \textit{Top-left}: Galactic longitude PM histogram of the bulge (red) and disk (blue) populations as well as the calculated peak separation. Vertical dashed line corresponds to our PM-selection cut at -2 mas yr$^{-1}$. \textit{Bottom-left}: Galactic latitude PM histogram of the same populations. \textit{Top-right}: Bulge-dominated (red) and disk-dominated (blue) regions of the CMD chosen for the PM analysis. \textit{Bottom-right}: PM vector point diagram of stars above the MSTO separated into the blue-plume branch (blue) and evolved bulge giant branch (red), all well-measured stars down to 21st mag are plotted in black. \label{fig:combinedPM}}
\end{figure*}

\section{Estimating Proper Motions in the Field} \label{para:3}
The stars above the main-sequence turnoff (MSTO) were used to calculate the mean PM of each population. The top-right panel of figure \ref{fig:combinedPM} shows the two regions used; the blue-plume (BP) and RGB regions respectively. The star counts in these two regions were kept similar to minimize the possible disk contaminants in the RGB sample following the procedure of \cite{Clarkson:2008aa} and C14a. We then calculated the mean PM of each population, finding $[\bar{\mu}_{\ell},\bar{\mu}_b]\approx(0.0, -0.3)$ mas yr$^{-1}$ with a dispersion of $[\sigma_{\ell}, \sigma_b] \approx (3.10, 3.01)$ mas yr$^{-1}$ for the RGB and $[\bar{\mu}_{\ell}, \bar{\mu}_b] \approx (3.5, -0.5)$ mas yr$^{-1}$ with dispersion $[\sigma_l, \sigma_b] \approx (3.20, 2.14)$ mas yr$^{-1}$ for the BP.\\
\indent Following this, we adopt a cut at $\mu_l \leq$ -2.0 mas yr$^{-1}$ to exclude $\sim$85\% of disk stars, while keeping $X\sim35\%$ of total bulge population stars. A more accurate estimation of $X$ can be made by integrating the Gaussian fit to $\mu_l$ (top-left panel of Figure \ref{fig:combinedPM}), from negative infinity to the PM cut. The integral is of the form:
\begin{equation} \label{eq:3}
X = \frac{1}{\sqrt{2\pi} \sigma} \int_{-\infty}^{-2.0} \exp(\frac{-(x-\bar{\mu})^{2}}{2\sigma^{2}}) dx,
\end{equation}

where $\bar{\mu}$ is the mean PM and $\sigma$ is the dispersion of the distribution. A detailed estimation of $X$ using this approach is described in section \ref{sec:EfficiencyCut}. We chose the PM cut of -2.0 mas yr$^{-1}$ by performing a simple optimization in which several thresholds were tested (i.e. -3.0, -2.5, -2.0, -1.5, -1.0 mas yr$^{-1}$ respectively) to find a resulting sample that best represents the cleaned distribution while maintaining statistical significance ($\sim$20,000 stars passing the rejection criteria). This analysis also considered the effect of cutting based on the calculated PM dispersions for each population as described in section \ref{sec:Dispersion}. The cutting threshold value of -2.0 mas yr$^{-1}$ was also adopted by the previous studies of \cite{Clarkson:2008aa} and \cite{Calamida:2015aa}. Finally, the optimal cutting threshold is likely a function of latitude location near the Galactic plane, but a detailed analysis of this function is beyond the scope of this paper.

\subsection{\normalfont{Proper Motion Dispersion}} \label{sec:Dispersion}
The PM dispersions $\sigma_l$, $\sigma_b$ and their uncertainties were estimated using two techniques. First, each PM distribution was fit by a Gaussian using a $\chi^2$ minimization routine, and then a dispersion and error on the dispersion were calculated from the resulting fit. This method accurately describes the distribution, with non-Gaussianity at most $\leq 10\%$ integrated across the distribution. The second method takes a more direct approach, following that of \cite{Kozlowski:2006aa} (which is based on \cite{Spaenhauer:1992aa} and references therein):
\begin{equation}
\sigma^{2} = \frac{1}{(n - 1)} \sum_{i=1}^{n} (\mu_{i} - \bar{\mu})^{2} - \frac{1}{n} \sum_{i=1}^{n} \xi_{i}^{2},
\end{equation}

where $\mu_{i}$ are the individual PMs and $\xi_{i}$ is the PM error (per coordinate) for the sample of $n$ stars with mean PM $\bar{\mu}$. The error in PM dispersion is then:
\begin{equation}
\xi_{\sigma} = \biggr(\frac{\sigma^{2}}{2n} + \frac{1}{12n^{2} \sigma^{2}} \sum_{i=1}^{n} \xi_{i}^{4}\biggr)^{\frac{1}{2}},
\end{equation}

The error in PM dispersion relies on the finite size of the sample $n$ and individual PM uncertainties. From these equations, we calculate PM dispersions ($\sigma_l \pm \xi_l$, $\sigma_b \pm \xi_b$) for the bulge-only population, disk-only population, and mixed bulge+disk population down to I $\approx$ 24. Both methods of calculation are in agreement, and we adopt the former method due to 15\% smaller PM errors on average across the entire magnitude range.\\

\begin{figure*}
\figurenum{4}
\plotone{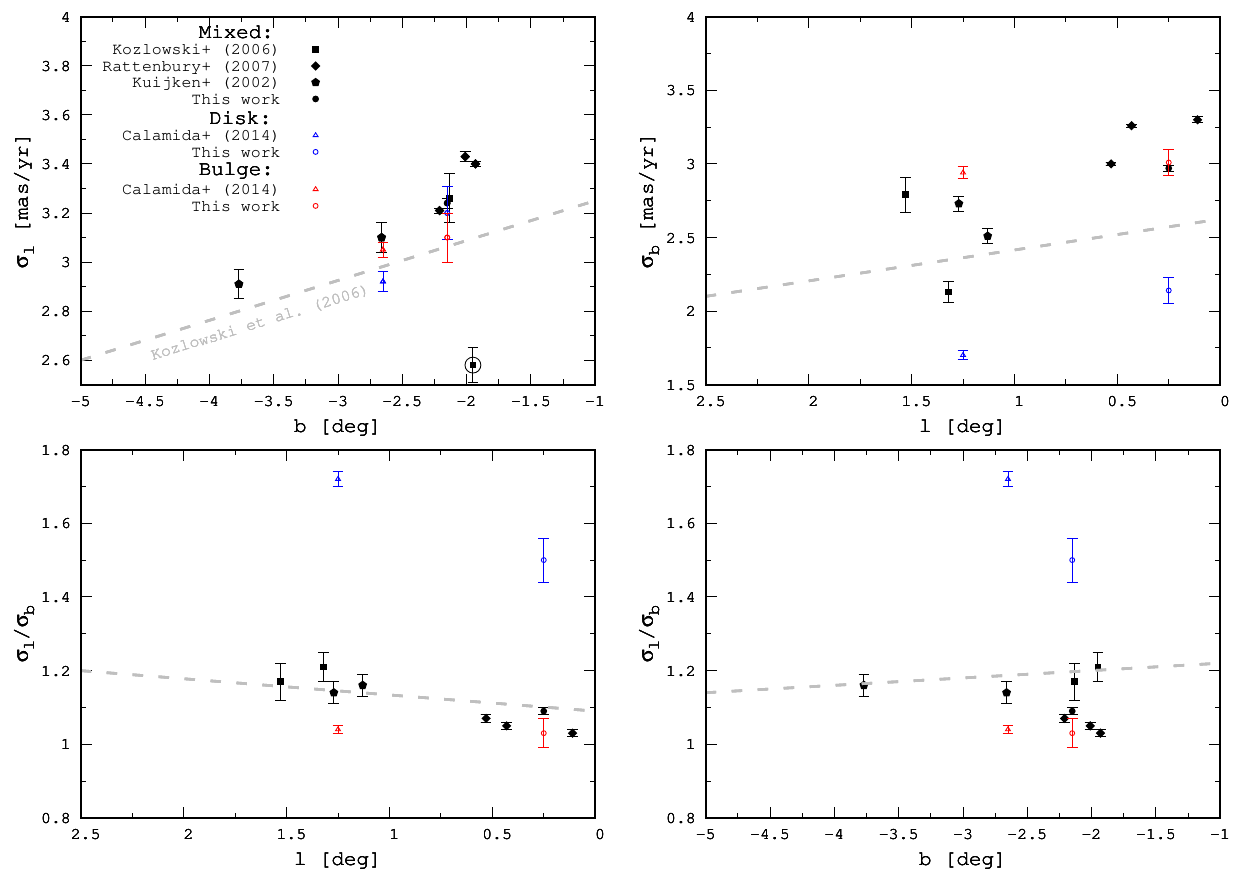}
\caption{\scriptsize Proper motion dispersions and anisotropy ratios for the sight-lines reported in Table \ref{table:2}. Black datapoints correspond to the `mixed' population. Blue and red datapoints correspond to the disk and bulge components as measured by \cite{Calamida:2014aa} and this work. The dashed line shows the linear regression fits from \cite{Kozlowski:2006aa}, and the circled datapoint in the top-left panel is the outlier, 97-BLG-41. \label{fig:table2plots}}
\end{figure*}

\indent Table \ref{table:2} and Figure \ref{fig:table2plots} show a comparison of these results to those of \cite{Kuijken:2002aa} for Sgr-I and BW and \cite{Calamida:2014aa} for the SWEEPS field. We also include prior well-studied bulge fields from \cite{rattenbury:2007aa} and \cite{Kozlowski:2006aa} that lie within $\sim1.5\degree$ of the Stanek field. The PM dispersion in the Stanek field is marginally larger in the longitude direction than most other bulge fields observed, which is to be expected for the nearest sight-line to the Galactic plane. The increasing contamination of disk stars at lower latitudes leads to a further spread in the longitudinal dispersion. Additionally, \cite{Kozlowski:2006aa} report a weak, but measurable gradient in $\sigma_{l}(b)$ and $\sigma_{b}(l)$ that increases with decreasing Galactic latitude and longitude. We also find evidence of this weak gradient in our Figure \ref{fig:table2plots} comparison (top panels). This weak gradient is also apparent in recent Gaia DR2 PM measurements \citep{brown2018gaia}.\\
\indent Further, the OGLE 97-BLG-41 field is the only sight-line with a lower latitude that has a significantly smaller dispersion in both $\sigma_l, \sigma_b$. This field is specifically pointed out by the authors as being intriguing in that it has the lowest measured dispersion, while being the lowest latitude field of their study. It turns out that the next nearest field in the \cite{Kozlowski:2006aa} study that meets our Table \ref{table:2} selection criteria, OGLE 98-BLG-6, is also somewhat of an outlier. While both fields are relatively nearby one another, they have the largest difference in dispersion amongst the 35 sight-lines in their sample. One explanation for this may be the increased extinction in these sight-lines. From \cite{nataf:2013aa}, the extinction in the 97-BLG-47 field is $A_{I} = 2.031$ and the extinction in the 98-BLG-6 field is $A_{I} = 1.560$.  The circled data point in the upper-left panel of Figure \ref{fig:table2plots} shows the 97-BLG-41 outlier. The authors decide not to use this data point in their linear regression fit for $\sigma_{l}(b)$.\\
\indent Finally, it is worth noting the Kozlowski dispersion measurements reported were performed on the mixed bulge+disk population in all sight-lines. Their sample was limited to the magnitude range $18.0 < I_{F814W} < 21.5$, that is dominated by bulge MS stars near the turn-off, but will undoubtedly still be contaminated by disk stars. The contamination should become most severe in their lowest latitude fields, which would seem at odds with their findings for the two fields described above. It is difficult to determine whether our dispersion results give further evidence for a subtle gradient in the rotational velocity of the bulge, as some of the previous studies have suggested.

\begin{deluxetable*}{lcccccccccc}
\deluxetablecaption{Proper Motion Dispersions Along Sight-lines \label{table:2}}
\tablecolumns{10}
\setlength{\tabcolsep}{3.5pt}
\tablenum{2}
\tablehead{
\colhead{{\hspace{-5mm}Field Name}} &
\colhead{Bulge/Disk} & \colhead{l} & \colhead{b} & \colhead{$\sigma_l$} & \colhead{$\sigma_b$} & \colhead{$\sigma_{l} / \sigma_{b}$} & \colhead{$\Delta$t} & \colhead{$N$} & \colhead{$^{*}A_{I}$} & \colhead{Reference}\\
\colhead{} &
\colhead{} & \colhead{[deg]} & \colhead{[deg]} & \colhead{[mas yr$^{-1}$]} & \colhead{[mas yr$^{-1}$]} & \colhead{} & \colhead{[year]} & \colhead{} & \colhead{}
}
\startdata
WFC3 Stanek & Both & 0.25 & -2.15 & 3.24 $\pm$ 0.02 & 2.97 $\pm$ 0.02 & $1.09 \pm 0.01$ & 2.00 & 10704 & 1.28 & This work\\
-- & Disk & -- & -- & 3.20 $\pm$ 0.11 & 2.14 $\pm$ 0.09 & $1.50 \pm 0.06$ & -- & 429 & -- & --\\
-- & Bulge & -- & -- & 3.10 $\pm$ 0.10 & 3.01 $\pm$ 0.09 & $1.03 \pm 0.04$ & -- & 487 & -- & --\\
ACS SWEEPS & Disk & 1.25 & -2.65 & 2.92 $\pm$ 0.04 & 1.70 $\pm$ 0.03 & $1.72 \pm 0.02$ & 9.25 & 2500 & 1.00 & C14b (\citeyear{Calamida:2014aa})\\
-- & Bulge & -- & -- & 3.05 $\pm$ 0.03 & 2.94 $\pm$ 0.04 & $1.04 \pm 0.01$ & -- & 2500 & -- & --\\
WFPC2 BW & Both & 1.13 & -3.77 & 2.91 $\pm$ 0.06 & 2.51 $\pm$ 0.05 & $1.16 \pm 0.03$ & 5.24 & 1076 & 0.76 & KR02 (\citeyear{Kuijken:2002aa})\\
WFPC2 Sgr-I & Both & 1.27 & -2.66 & 3.10 $\pm$ 0.06 & 2.73 $\pm$ 0.05 & $1.14 \pm 0.03$ & -- & 1388 & 0.96 & --\\
97-BLG-41 & Both & 1.32 & -1.95 & 2.58 $\pm$ 0.07 & 2.13 $\pm$ 0.07 & $1.21 \pm 0.04$ & 5.15 & 612 & 2.03 & K06 (\citeyear{Kozlowski:2006aa})\\
98-BLG-6 & Both & 1.53 & -2.13 & 3.26 $\pm$ 0.10 & 2.79 $\pm$ 0.12 & $1.17 \pm 0.05$ & 4.25 & 670 & 1.56 & --\\
OGLE-II 3 & Both & 0.11 & -1.93 & 3.40 $\pm$ 0.01 & 3.30 $\pm$ 0.02 & $1.03 \pm 0.01$ & 3.91 & 26763 & 1.63 & R07 (\citeyear{rattenbury:2007aa})\\
OGLE-II 4 & Both & 0.43 & -2.01 & 3.43 $\pm$ 0.02 & 3.26 $\pm$ 0.01 & $1.05 \pm 0.01$ & -- & 26382 & 1.49 & --\\
OGLE-II 39 & Both & 0.53 & -2.21 & 3.21 $\pm$ 0.01 & 3.00 $\pm$ 0.01 & $1.07 \pm 0.01$ & -- & 24820 & 1.48 & --\\
\enddata
\tablenotetext{}{\footnotesize{\textbf{Notes}. Dispersions for SWEEPS given by Calamida via private comm.\\
$^{*}$ Extinction from \cite{nataf:2013aa}}}
\end{deluxetable*}

\subsection{\normalfont{Proper-Motion Cutting Efficiency}} \label{sec:EfficiencyCut}
\begin{figure}
\figurenum{5}
\epsscale{1.2}
\plotone{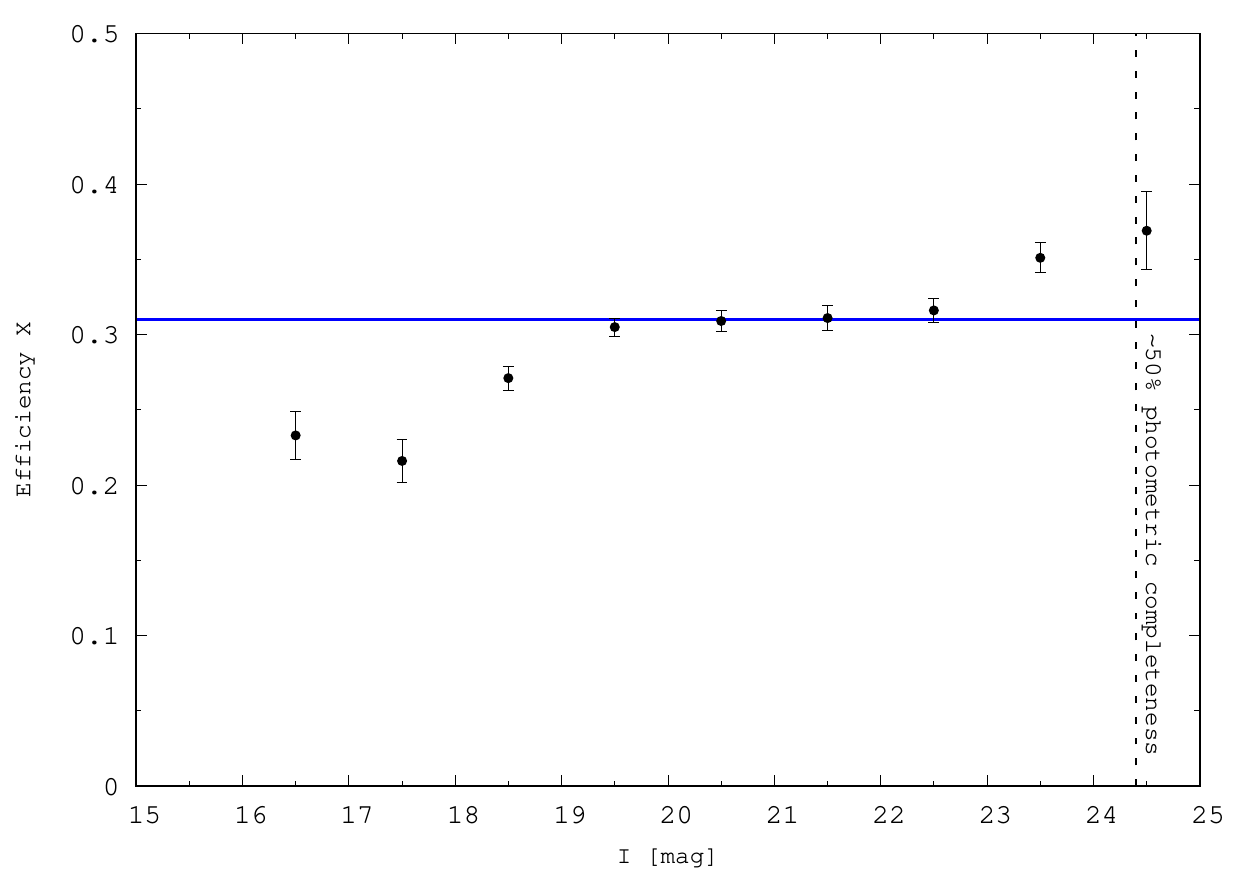}
\caption{\scriptsize{PM cutting efficiency as a function of I magnitude. Blue horizontal line represents the efficiency measured from the tracer region above the MSTO (Figure 3).} \label{fig:efficiencies}}
\end{figure}

In order to scale the cleaned bulge LF up to a `full' bulge LF, it is important to estimate a correction for the efficiency of the PM cut that is being made. \cite{Calamida:2015aa} do not report any correction for their SWEEPS PM-cut efficiency, however P19 subsequently calculated an efficiency of the (bulge-only) PM-cut of 34 percent based on an estimate similar to that described in equation \ref{eq:3} of this paper. The value of $X\sim0.34$ is used to scale the entire cleaned SWEEPS LF up to the `full bulge' LF. For the Stanek LF, we calculated an efficiency factor in two ways. In the first method, we applied equation \ref{eq:3} to the RGB population stars from the tracer region described earlier in this section. We find an efficiency factor $X\sim0.32$. In the second method, we applied the same equation to the fully mixed (bulge + disk) PM-distribution as a function of I magnitude from approximately the red clump down to the 50\% completeness threshold of I $\approx$ 24.4. This results in an efficiency curve that covers each relevant magnitude bin. Figure \ref{fig:efficiencies} shows the PM-cutting efficiency curve for $X$, which increases from $\sim$0.23 at the brightest magnitudes to $\sim$0.36 at faint magnitudes. This curve flattens at $X$ $\sim$ 0.31 between intermediate magnitudes $19.5-23$ corresponding to most of the observed MS, and is due to the two populations being fully mixed and indistinguishable from one another. The efficiency covering this flat region is consistent with the value calculated via method one. Method one will clearly under-correct the LF at brighter magnitudes and over-correct at the faint magnitudes, by upwards of $\sim$8\%. To avoid introducing any additional inaccuracies, we adopt the second method for our PM-efficiency scaling.\\
\indent Table \ref{table:3} gives the details of each bin and the convolved Gaussian sigma from combining the PM distribution and the individual PM-error distributions. Again, the propagated errors are smaller by $\sim 15$\% using equation \ref{eq:3}. The fit to the distribution broadens as expected and increases at fainter bins, while the contribution from the PM-error distribution has a measurable, but small effect on most of the full bulge LF. The residual disk star contamination increases at fainter magnitude bins, which is nearly impossible to accurately calculate. However, using estimates from \cite{Gennaro:2015aa}, we place a rough constraint of $0.5-3.0$\% residual contamination that spans the range I = 16 - 25. Lastly, standard error propagation techniques are used for the values reported in table \ref{table:3}, and associated error for the convolved Gaussian ${\sigma_l + \xi_{l}}$ is computed by summing the prior Gaussian errors in quadrature.

\begin{deluxetable}{lccccccc}
\tablecaption{Proper-Motion Cutting Efficiency \label{table:3}}
\tablecolumns{6}
\setlength{\tabcolsep}{2.5pt}
\tablenum{3}
\tablewidth{0pt}
\tablehead{
\colhead{Bin} &
\colhead{$\langle \mu_{l} \rangle$} & \colhead{$\sigma_l$} & \colhead{${\sigma_l + \xi_{l}}$} & \colhead{$X$}\\
\colhead{[mag]} & \colhead{[mas yr$^{-1}$]} &
\colhead{[mas yr$^{-1}$]} & \colhead{[mas yr$^{-1}$]} & \colhead{}
}
\startdata
16 -17 & {$0.73 \pm 0.12$} & {$3.749 \pm 0.119$} & {$3.751 \pm 0.119$} & {$0.233 \pm 0.016$}\\
17 - 18 & {$1.05 \pm 0.11$} & {$3.875 \pm 0.110$} & {$3.877 \pm 0.110$} & {$0.216 \pm 0.014$}\\
18 - 19 & {$0.27 \pm 0.06$} & {$3.726 \pm 0.058$} & {$3.728 \pm 0.058$} & {$0.271 \pm 0.008$}\\
19 - 20 & {$-0.19 \pm 0.04$} & {$3.550 \pm 0.042$} & {$3.555 \pm 0.042$} & {$0.305 \pm 0.006$}\\
20 - 21 & {$-0.24 \pm 0.04$} & {$3.517 \pm 0.044$} & {$3.527 \pm 0.044$} & {$0.309 \pm 0.007$}\\
21 - 22 & {$-0.30 \pm 0.05$} & {$3.454 \pm 0.050$} & {$3.479 \pm 0.051$} & {$0.311\pm 0.008$}\\
22 - 23 & {$-0.33 \pm 0.05$} & {$3.482 \pm 0.038$} & {$3.568 \pm 0.043$} & {$0.316 \pm 0.008$}\\
23 - 24 & {$-0.56 \pm 0.07$} & {$3.767 \pm 0.064$} & {$3.932 \pm 0.071$} & {$0.351 \pm 0.010$}\\
24 - 25 & {$-0.28 \pm 0.26$} & {$5.123 \pm 0.262$} & {$6.122 \pm 0.402$} & {$0.358 \pm 0.026$}\\
\enddata
\tablenotetext{}{\footnotesize{\textbf{Notes}. $\sigma_{l} + \xi_{l}$ is the convolution of both Gaussian distributions, with associated uncertainty estimated by summing the Gaussian errors in quadrature. A larger $X$ value corresponds to smaller relative scaling.}}
\end{deluxetable}

\subsection{ \normalfont{Anisotropy Ratio}} \label{sec:anisotropy}
\indent The anisotropy ratio was also calculated for the Stanek field and found to be $\sigma_{\ell} / \sigma_{b}=1.03$ for Bulge-only, $\sigma_{\ell} / \sigma_{b}=1.50$ for disk-only, and $\sigma_{\ell} / \sigma_{b}=1.09$ for the mixed populations respectively. All are accurately measured despite the conservative number of proxy stars used. \cite{Spaenhauer:1992aa, Kuijken:2002aa, Kozlowski:2006aa} and others suggest an anisotropy ratio measurably larger than 1 due to the rotating bulge-bar component. Our results show a pure-bulge anisotropy ratio closer to 1, which is in contrast with the prior studies. However, our `mixed' population anisotropy measurement shows some evidence of rotation, but is likely due to the additional contribution of the foreground disk population. We caution that it may be difficult to draw any conclusive trends between the sight-lines due to a majority of the prior fields only having measurements of the mixed bulge+disk populations rather than any pure component.\\
\indent While this discrepency may not seem severe, as \cite{Kuijken:2002aa, Kozlowski:2006aa} have stated, there are clearly measurable differences between our `Bulge' only dispersions and `Both' mixed dispersions in Table \ref{table:2}. Recall that at bulge distances, a PM of $\sim3.0$ mas yr$^{-1}$ corresponds to a transverse velocity of $\sim115$ km s$^{-1}$. The smaller anisotropy measured for the Stanek field `Bulge' and/or `Both' populations is in agreement with the trend reported in \cite{Kozlowski:2006aa}, showing a decrease in anisotropy nearer to the minor axis (i.e. $\ell = 0$). Our measured dependence of $\sigma_{l}/\sigma_{b}$ on $b$ does not agree as well with the fields in these prior studies, however it is marginally consistent with the observed $\sigma_{l}/\sigma_{b}$ scatter.\\
\indent We remind the reader that all of these fields reside in a relatively small area of the bulge, within $\sim1.5\degree$ of the minor axis and within $\sim3\degree$ of the plane. Our measured dispersion gradient also agrees with the recent study of \cite{clarke:2019aa} who analyzed PM data from the Vista Variables in the Via Lactea (VVV) and the \textit{Gaia} DR2 survey. It's important to emphasize the dependence of $\sigma_{\ell}$ on $b$ and $\sigma_{b}$ on $\ell$ and is apparent when comparing dispersed fields against the lowest longitude field in the set by far (Stanek at $l=0.25$). Explanations of this dependance and other anisotropy descriptions that go further into detail are given in \cite{Clarkson:2008aa} and references therein.

\subsection{\normalfont{Rotation Curve}} \label{sec:RotCurve}
\begin{figure}
\figurenum{6}
\epsscale{1.2}
\plotone{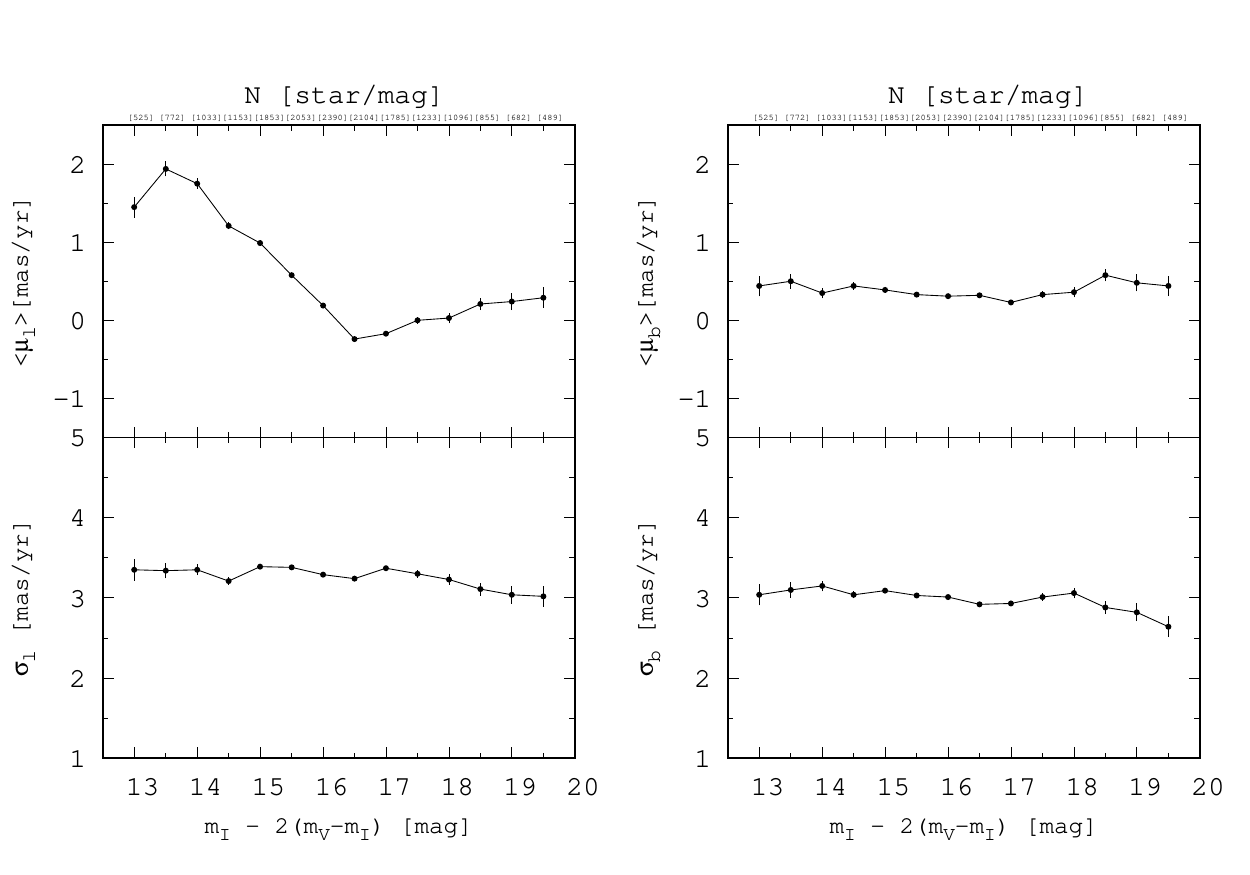}
\caption{\scriptsize{\textit{Left}: Galactic longitude mean PMs and dispersions of Stanek field stars as a function of the distance indicator. \textit{Right}: Galactic latitude component of the motions.}\label{fig:RotCurve}}
\end{figure}

Following \cite{Kuijken:2002aa} and \cite{Kozlowski:2006aa}, a crude distance modulus can be calculated by removing the slope of the CMD and cutting a cross-section along the new de-colored MS. This results in a simple relative distance indicator for each star in the set, which follows the form:\\
\begin{equation}
M^* = m_{I} - 2(m_{V} - m_{I}),
\end{equation} \\
Stars are binned by distance indicator and plotted against their mean PMs and PM dispersions, which are shown in Figure \ref{fig:RotCurve}. As expected, the kinematics along the line of sight describe the near-side foreground disk stars, transitioning through the bulge population and very likely beyond the bulge to the far-side stars. This is detailing a rotation curve for the Milky Way along this sight-line in terms of the relative PM and their dispersions, with limited resolution at the faintest magnitudes corresponding to backside disk star populations. The amount of contamination by disk star rotation is not large enough to significantly influence the mean velocity for all stars in a given distance indicator bin.\\
\indent For detailed descriptions of the velocity profile and kinematics, we refer the reader to \cite{zhao:1994aa}, \cite{izumiura:1995aa}, \cite{zhao:1995aa}, and more recently \cite{clarkson:2018aa}.

\section{Bulge Star Counts} \label{para:4}
\begin{figure}[!h]
\figurenum{7}
\includegraphics[scale=0.65]{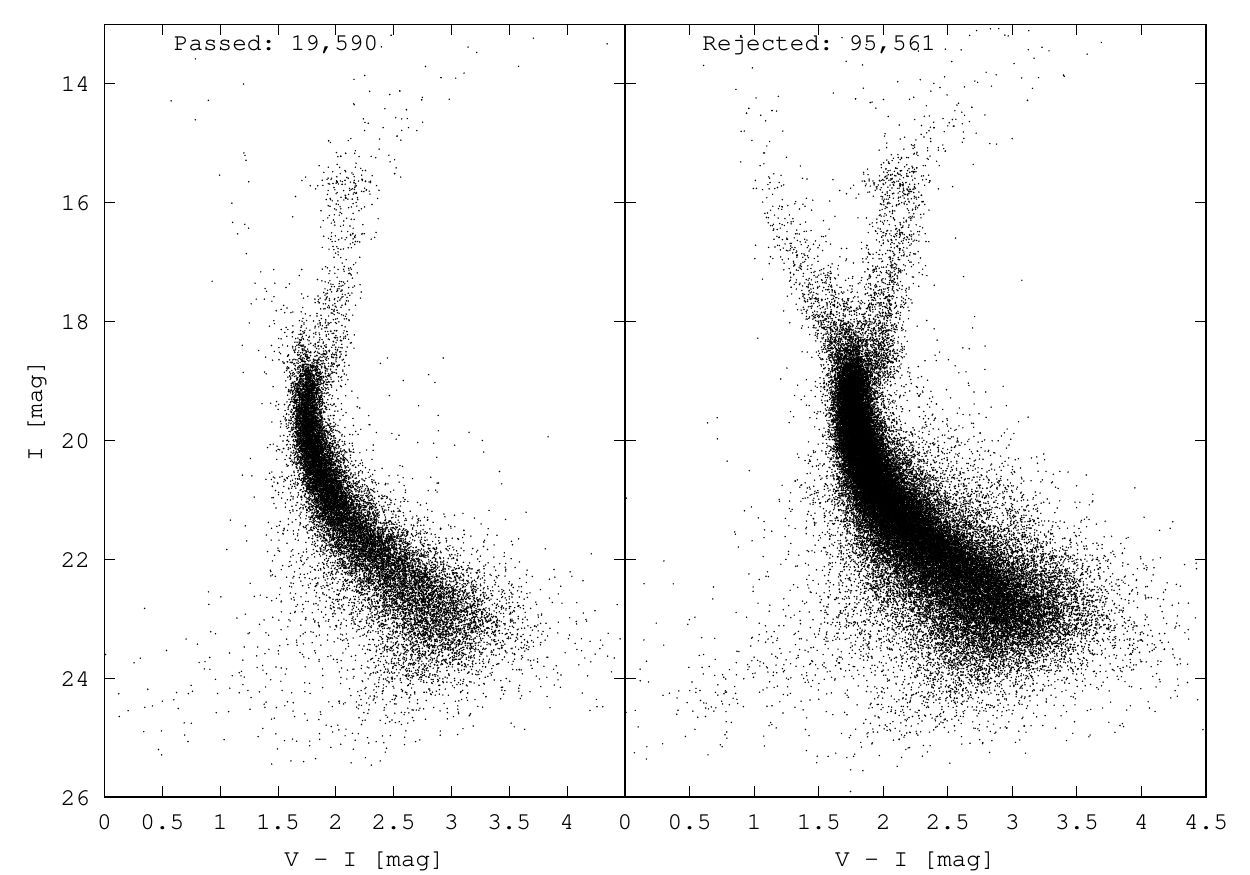}
\centering
\caption{\scriptsize \textit{Left}: CMD of stars that passed the PM cut. \textit{Right}: CMD of rejected stars from the cut. \label{fig:cleanedCMD}}
\end{figure}

The completeness corrections and PM-cutting efficiency leads to a proper accounting of all bulge sources along this sight-line. Figure \ref{fig:cleanedCMD} shows the cleaned CMD after PM cuts have been applied (left panel), along with stars that are rejected as a result of the cut (right panel). Such a large rejection threshold is needed to ensure the final dataset is sufficiently cleaned of foreground star contamination. The cleaned CMD of the Stanek field shows some evidence of a population of younger bulge BSS just brighter and bluer than the MSTO. This would seem to verify the results of \cite{Clarkson:2011aa}, although we do not conduct a further analysis in the current paper. There is also a small amount of residual contamination from foreground stars that remain in the cleaned dataset. These sources have high enough (e.g. $<-2$ mas yr$^{-1}$) PM to pass the cut; they are likely bright foreground disk stars with sporadically high PM, or large bright stars counter-rotating on the far side of the bulge/disk. Additionally, there is evidence of the WD cooling sequence in the bulge previously detected by C14a that reside in a similar location on the cleaned CMD ($0 < V-I < 1.5$ and $23 < I < 25$).

\begin{figure}
\figurenum{8}
\epsscale{1.2}
\plotone{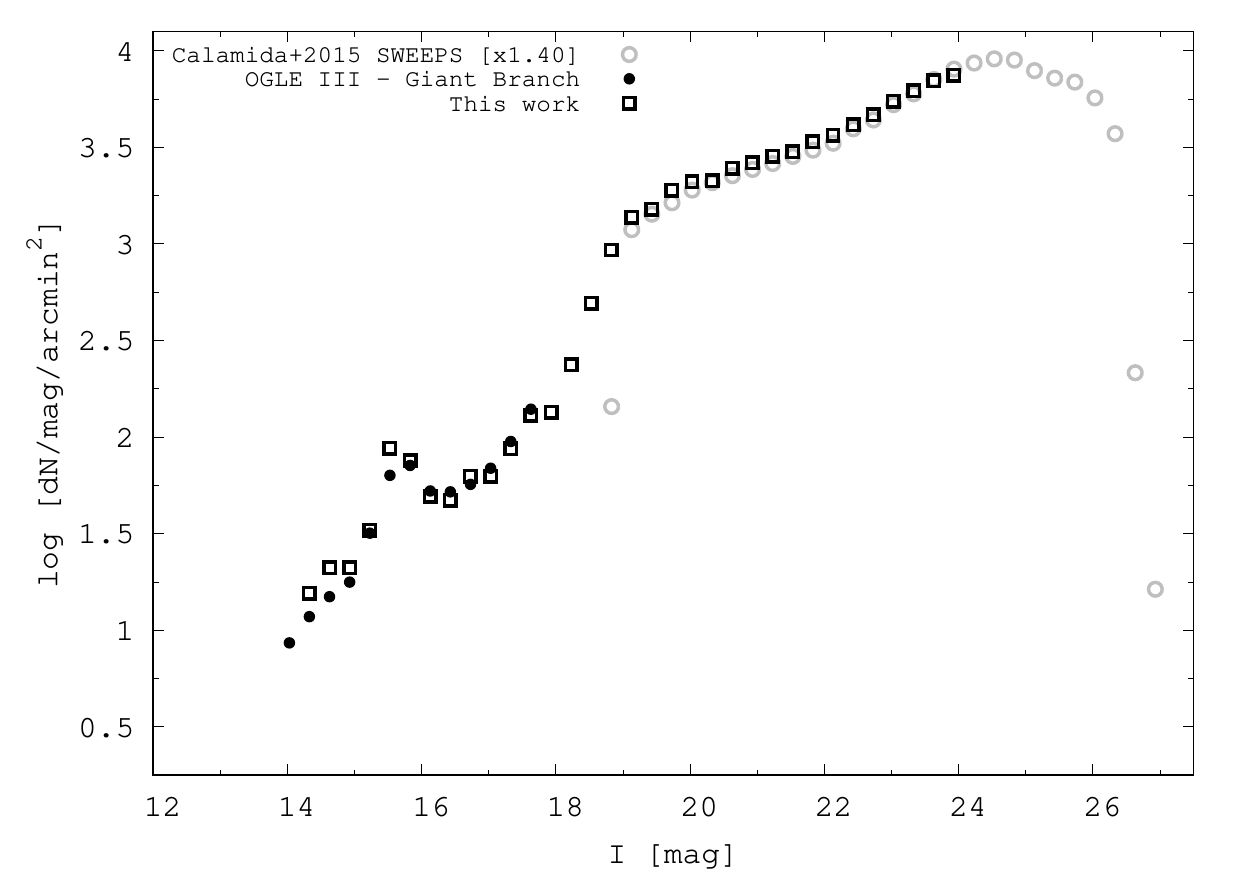}
\caption{\scriptsize Stanek field bulge star counts compared to \cite{Calamida:2015aa} SWEEPS field bulge star counts (grey). RGB stars from OGLE-III \citep{szymanski:2011aa} in the Stanek field are plotted as solid black points. The SWEEPS counts are scaled by $\sim$40\% to match the larger surface density of the Stanek field. \label{fig:finalLF}}
\end{figure}

\indent The bulge-only star counts are presented in Figure \ref{fig:finalLF}, after being corrected for photometric completeness and PM-cutting efficiency. Data with low completeness ($\lesssim 50\%$ at $I \gtrsim 24$) are not plotted. Star counts from the OGLE-III \citep{szymanski:2011aa} fields overlapping the HST Stanek coverage are shown covering primarily the RC at $I \sim 15.7$. The OGLE-III stars have not been completeness-corrected. The very deep SWEEPS field bulge star counts of \cite{Calamida:2015aa} are shown in Figure \ref{fig:finalLF} as well. The SWEEPS star counts are found to be approximately 40\% less than Stanek integrated over the magnitude range I = 19.5-23.5. The increased surface density between these fields is expected, but slightly larger than values estimated by \cite{wegg:2013aa} who show projections of the fiducial density measurements of RC stars identified in DR1 of the VVV survey \citep{saito:2012aa}. The $\sim7\%$ larger than expected surface density we find is due to several factors. A higher residual disk star contamination in our pure-bulge sample is likely caused by a marginally lower cleaning efficiency, which can be seen as a difference between the disk population $\sigma_{l}$ reported on Table \ref{table:2}. As described earlier, the nearer location of the Stanek field to $\ell = 0$, $b = 0$ cause a more severe contamination. Recall the center of the Stanek field is $\sim$$1.0\degree$ closer to the minor axis and $\sim$$0.5\degree$ closer to the major axis relative to the center of SWEEPS.

\subsection{ \normalfont{RC/MS Ratio}} \label{sec:rc-ms}
To confirm there is no intrinsic difference in the bulge-only population of evolved RGB stars and MS stars between Stanek and SWEEPS, we measured the ratio of the surface densities for the RC and MS range in the two fields. For the RC measurement in the Stanek field, we used the \texttt{BLG101.3.map} catalog from OGLE-III. This star list is nearest to the Stanek HST field with significant overlap. We measure the centroid of the RC in this field to be $I_{rc}=15.721$ and $(V-I)_{rc}=2.099$, and from here we measure the RC star count in the window by integrating over the range $I=15.571-15.871$. The MS star count is measured by integrating the HST WFC3 star counts over $I=21-23$. Finally, we measure RC/MS$_{Stanek} = 0.020$ for the raw counts.\\
\indent Similarly, for the SWEEPS field we used \texttt{BLG104.5.map} from OGLE-III for the RC measurements. The RC centroid was measured to be $I_{rc}=15.376$ and $(V-I)_{rc}=1.843$. To account for the difference in the I magnitude location of the RC between the fields ($\Delta I=0.345$), the MS magnitude range integrated over for SWEEPS was $I=20.65-22.65$. We measure RC/MS$_{Sweeps}$ = 0.021 for the raw counts. Our results show that the ratio's for both fields are consistent and confirm that there is indeed no significant intrinsic variation in population types amongst these two bulge sight-lines.

\subsection{\normalfont{Near-IR Star Counts}} \label{para:5}
The version 2 science products from the WFC3 Bulge Treasury Program \citep{brown:2009aa} were used to analyze the J(F110W) and H(F160W) photometry, astrometry, completeness, and surface density. The methods described in Section \ref{sec:2} were used for these redder data, with the exception of the photometric calibration. The photometry given by the version 2 data products are in the STMAG system, thus we converted these magnitudes to VEGAMAG using the zero-points from \cite{deustua:2017aa}. The FoV for the WFC3/IR camera is smaller than UVIS, at a scale of 136$''$x123$''$, so additional offsets were performed to provide the IR camera full coverage of the UVIS area (162$''$x162$''$). The total exposure time for the J-band data was 1255s, and 1638s for the H-band data.\\
\indent The right panel of Figure \ref{fig:LF+Ratio} shows the H-band LF for bulge stars in the Stanek field along with three simulated LF's (normalized to dN/mag/arcmin$^{2}$ and transformed to VEGAMAG) from the models. The slope of the LF below the MSTO is steeper than the I-band results as expected for a bulge with a significant amount of low-mass dwarfs. Of the three population synthesis models, GalMod most closely predicts the surface density in this near-IR band, however the higher low-mass cutoff of the IMF precludes the model from probing the dimmest population stars. From this, we are unable to deduce whether the modeled LF `turns-up' (like Galaxia), or `turns-down' (like Besan\c{c}on) at the faintest magnitudes. A future update to the GalMod software will include the MIST stellar tracks \citep{choi:2016aa, dotter:2016aa}, which will allow for much deeper CMD's and LF's (S. Pasetto private comm). Lastly, Table \ref{table:4} shows comparisons between relevant parameters for each population synthesis model. The major difference again is the low-mass contribution in the IMF's and the bar-angle.

\subsection{ \normalfont{Comparison with Bulge Population Synthesis Models}} \label{sec:model-compare}
As a basis for estimating WFIRST exoplanet yields, population synthesis models are used to generate stellar surface densities and microlensing optical depths, which in turn can be used to estimate microlensing event rates. Based on the event rate, a detection efficiency for microlensing planets can then be derived while placing limits (via the models) on expected planet yields. The better a model can accurately describe surface densities in very crowded and highly extincted fields, the more accurate the estimations of planet detection efficiencies can be (P19). The P19 simulation study is a very detailed project that has estimated the planet detection efficiency and expected exoplanet yields for the current Cycle 7 WFIRST design. P19 compared only one Galactic model in their study and were subsequently required to make adjustments to the yields in order to match observations after discrepancies in the model were considered. We expand on this part of the P19 study by comparing several popular Galactic models to the Stanek field observations. \\
\indent Firstly, the models analyzed here are all publicly available either by direct access via web interface or download and compile. Second, the models all generally describe the central bulge as a boxy triaxial bar shape \citep{Dwek:1995aa} with two models using a bulge-bar angle of $\sim$$12-13\degree$ with respect to the Sun-Galactic line and the third model assumes an angle of $\sim$$28\degree$. They all use relatively similar scale lengths, each incorporates disk and bulge kinematics, and they include some form of prescription for the warp and flare of the thin disk and bulge respectively. The built-in photometric system chosen for each model is the Johnson Cousins (UBVRIJHKL) system. As we show in the following sub-sections, each model has its strengths and weaknesses, therefore it is a worthy exercise to perform the comparison of these sophisticated models to a set of real data. Lastly, It is important to keep in mind a caveat; the Galactic bulge-bar is arguably the most complex MW component and it is very likely impossible for any model of our galaxy to perfectly simulate all of the complexities that lie within such a complicated dynamical system.

\begin{figure}
\figurenum{9}
\plotone{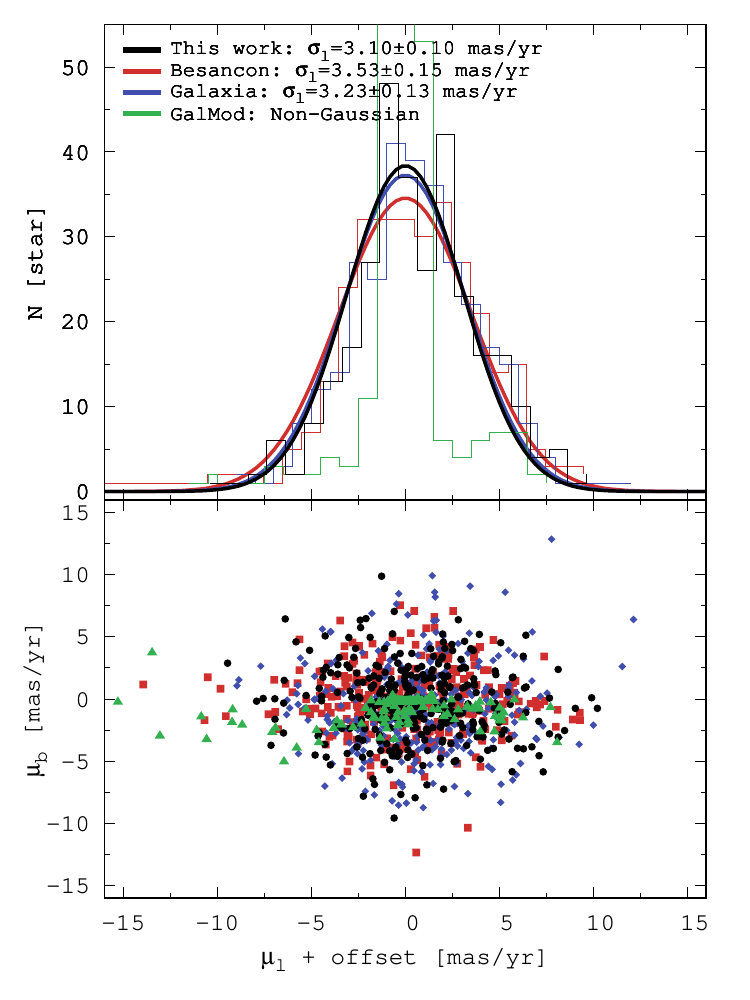}
\caption{\scriptsize \textit{Top}: PM histogram for the observed Stanek field (black curve) mean longitudinal component, $\mu_l$, compared to the three population synthesis models. \textit{Bottom}: Vector point diagram of bulge proxy stars selected from the RGB in each sample. Selection size is $\sim400$ stars for each sample. \label{fig:pm-diag}}
\end{figure}

\subsubsection{ \normalfont{Besan\c{c}on}} \label{sec:besancon}
Version 1106 of the Besan\c{c}on Galactic Model \citep{robin:2003aa, robin:2012aa} (hereafter BGM1106) was used by P19 for their analysis and subsequent microlensing event rate and detection efficiency estimation. They give a detailed description of BGM1106 in \cite{Penny:2013aa}, so the details given here will be limited to the differences between the P19 model and the current model used in this paper (version 1612, hereafter BGM1612\footnote[9]{\href{https://model.obs-besancon.fr/index.php}{\scriptsize https://model.obs-besancon.fr/index.php}}). A full description of the uncertainties in BGM1106 can be found in section 6.2 of P19.\\
\indent BGM1612 uses a slightly larger bulge-bar angle of $12.9\degree$, a thin disk density law \citep{einasto:1979aa} with an 8\% smaller scale length of 2.17 kpc for the old stars and the same scale length of 5 kpc for the young stars. There is a central hole in the disk with a scale length of 1.33 kpc for old stars and 3 kpc for young stars, which is virtually unchanged between versions. P19 measure the BGM1106 kinematics toward the SWEEPS field, and find PM dispersions roughly similar to the detailed PM study of \cite{Clarkson:2008aa}, with the exception of the Galactic longitude dispersion $\sigma_{l}$ for the red-bulge population stars. The authors find the dispersion from the simulation to be larger than the observed value by a factor of $1.73 \pm 0.12$, which leads to microlensing event timescales that are too short. The too-fast kinematics in the model have since been corrected in the current version, and we find only a marginal increase in the logitudinal PM dispersion between the current model and observations. As shown in Figure \ref{fig:pm-diag}, BGM1612 has a larger dispersion by a factor of $1.14 \pm 0.05$. Note the circular velocity (at the distance of the Sun) that the current version uses is $V_{LSR}$ = 244.6 km s$^{-1}$. Further, the PM's reported by BGM1612 and the subsequent models are in an absolute frame, whereas the Stanek HST measurements are made in an arbitrary frame since there is no absolute reference to measure the motions against. We apply a simple offset to the models bulge and disk proxy populations to correct for this.\\
\indent The bulge-bar angle of 12.9$\degree$ that the model assumes is significantly smaller than some prior observational studies \citep{stanek:1994aa, rattenbury:2007aa, cao:2013aa} who measure a bar angle $\sim25-45\degree$. P19 point out that a smaller bar angle leads to a larger spread of bulge stars along the line-of-sight. This ultimately results in larger Einstein radii and event timescales for bulge-bulge lensing and, to a lesser extent, bulge-disk lensing. The Einstein radius depends on the relative distance to the source and lens by:

\begin{equation}
r_{E} = \sqrt{\frac{4G}{c^{2}}M_{l}D_{l}(1-\frac{D_{l}}{D_{s}})},
\end{equation}

\noindent where $G$ and $c$ are the fundamental constants for gravity and speed of light, $M_{l}$ is the mass of the lens, and $D_{l}$ and $D_{s}$ are the distance to the lens and source. It is also clear with a smaller bar angle, the ratio of bulge lenses to disk lenses will be larger. \cite{robin:2012aa} attempt to reconcile this discrepancy by comparing different versions of their model with varying bar angles to determine if they get a more favored fit. They conclude that a larger bar angle gives a higher likelihood at locations further from the minor axis and close to the plane, but a lower likelihood nearer to the minor axis and higher latitude. Their results agree with the study of \cite{cabrera:2007aa, cabrera:2008aa} from an analysis of 2MASS data, however several authors have pointed out that the 2MASS data does not probe the RC population. The survey is much shallower and only measures the upper red giant population with a broader range in luminosity and much larger fraction of disk contamination. Additionally, \cite{simion:2017aa} presented a 3D description of the bar/bulge from the VVV survey and find a strong degeneracy between the bar angle and the RC absolute magnitude dispersion. This degeneracy may be what is causing the Besancon Galactic Model (BGM) to under-predict the bar angle.\\
\indent One final discrepancy between BGM and observations is the choice of bulge IMF the model uses. BGM1612 uses the Padova Isochrones \citep{marigo:2008aa,bressan:2012aa}, with a broken power-law, d$n$/d$m \propto m^{-\alpha}$ for the bulge population. An IMF slope of $\alpha = 0.5$ is used for the low-mass range $0.15M_{\sun} < m < 0.7M_{\sun}$ and $\alpha = 2.3$ for $m > 0.7M_{\sun}$. The low-mass cutoff of 0.15$M_\sun$ for the IMF is higher than that needed for the bulge population, as a major fraction of bulge stars will be very low mass (VLM) dwarfs \citep{Calamida:2015aa}. The low-mass IMF slope of $\alpha = 0.5$ is shallower than the BGM1106 model which used a slope of $\alpha = 1.0$, both of which are shallower than low-mass slopes obtained from observations. P19 fit a more reasonable slope from \cite{sumi:2011aa} to the model and found better agreement with the shape of the SWEEPS LF, but ultimately decided to keep the BGM1106 low-mass slope and correct for the inaccuracy in their further analysis. Adding VLM stars (e.g., $m < 0.15M_\sun$) to the model IMF will result in better agreement in surface densities at the dimmest magnitudes, especially in redder wavebands which WFIRST will utilize for its microlensing survey. It's worth noting that adding brown dwarfs (BD) will also clearly increase the surface density, however the mass function of BD in the bulge is quite uncertain \citep{sumi:2011aa, mroz:2017aa, wegg:2017aa}. Adding these VLM stars and some fraction of BD to the bulge IMF will result in increased optical depth and event rates per star for bulge-bulge lensing.

\begin{deluxetable*}{lccccccc}
\tablecaption{Bulge Parameters for Each Population Synthesis Model \label{table:4}}
\tablecolumns{6}
\setlength{\tabcolsep}{2.5pt}
\tablenum{4}
\tablewidth{0pt}
\tablehead{
\colhead{Reference} &
\colhead{Age} & \colhead{Fe/H} & \colhead{IMF} & \colhead{SFR} & \colhead{Bar-Angle}\\
\colhead{} & \colhead{[Gyr]} &
\colhead{[dex]} & \colhead{[$dn/dm \propto m^{-\alpha}$]} & \colhead{} & \colhead{[$\degree$]}
}
\startdata
Besan\c{c}on 1612 & {10.0} & {$0.00 \pm 0.40$} & {$\alpha=2.3, m \geq 0.7M_{\sun}$} & {single burst} & {12.9}\\
{} & {} & {} & { $\alpha=0.5, 0.7M_{\sun} > m \geq 0.15M_{\sun}$} & {} & {}\\
& {} & {} & {} & {} & {}\\
Galaxia 0.7.2 & {10.0} & {$0.00 \pm 0.40$} & {$\alpha=2.3, m \geq 0.7M_{\sun}$} & {single burst} & {12.9}\\
{} & {} & {} & { $\alpha=0.5, 0.7M_{\sun} > m \geq 0.15M_{\sun}$} & {} & {}\\
{} & {} & {} & { $\alpha=1.3, 0.15M_{\sun} > m \geq 0.07M_{\sun}$} & {} & {}\\
& {} & {} & {} & {} & {}\\
GalMod 18.19 & {6.0$-$12.0} & {$0.00^{+0.3}_{-0.4}$} & {$\alpha=2.3, m \geq 1.0M_{\sun}$} & {\cite{rosin:1933aa}} & {27.9}\\
 & {} & {} & {$\alpha=2.7, 1.0M_{\sun} > m \geq 0.5M_{\sun}$} & {} & {}\\
 & {} & {} & {$\alpha=1.8, 0.5M_{\sun} > m \geq 0.16M_{\sun}$} & {} & {}\\
\enddata
\end{deluxetable*}

\subsubsection{ \normalfont{Galaxia}} \label{sec:galaxia}
The Galaxia\footnote[10]{\href{http://galaxia.sourceforge.net/}{\scriptsize http://galaxia.sourceforge.net/}} version \texttt{0.7.2} code \citep{sharma:2011aa} largely implements the Besan\c{c}on model, but includes a wide variety of input parameters with more flexibility than that of BGM1612. Some adjustable parameters include the choice of an analytical or N-body seeded model from \cite{bullock:2005aa}, the ability to sub-sample the simulated data by a given fraction in order to reduce runtimes and file size, and no restriction to the size of a given catalogue (Besan\c{c}on has set a new maximum of two million stars per simulation as of May 2019). The model is run by \texttt{C++} compilation. As Galaxia is essentially an altered version of BGM with extended capabilities, we proceed further with an analysis of the three discrepancies which were described in the previous sub-section; the bulge IMF, the $\mu_{\ell}$ kinematics, and the bulge-bar angle, in order to investigate any differences in their significance within this modified software:
\begin{enumerate}
\item Like Besan\c{c}on, Galaxia uses the Padova isochrones for the bulge population, but importantly also includes isochrones from \cite{chabrier:2000aa} for the lower mass regime $0.07M_{\sun} < m < 0.15M_{\sun}$, below the BGM1612 cutoff. This results in better agreement with surface densities at the faintest magnitudes in both the I-band and the H-band LF's. Although Besan\c{c}on can model WD's, this version of Galaxia does not account for them. This ultimately has little to no effect on our interpretation of star counts as this population is very faint and does not overlap our observation range. Lastly, BDs are not being modeled in this code, which may affect the resulting surface densities at the faintest levels but likely much fainter than the magnitude range of interest.
\item The analysis of the Galaxia kinematics follows the same procedure described in Section \ref{sec:besancon}, with the focus of comparing the modeled $\sigma_{l}$ values with the Stanek field measurements. Figure \ref{fig:pm-diag} shows the PM diagram and histogram comparison. The Galaxia longitudinal dispersion $\sigma_l = 3.23 \pm 0.13$ is consistent with both the observed value from the Stanek HST measurements and the BGM1612 dispersion. The circular velocity that Galaxia uses is marginally lower than BGM1612, at V$_{c}$ = 224.8 km s$^{-1}$.
\item Galaxia adopts the same bulge-bar angle of $\sim$13$\degree$ that BGM uses, which again is smaller than the results found by numerous observational studies in the past. Both models incorporate warp and flare of the thin/thick disk, derived from \cite{robin:2003aa}. As stated above, the Galaxia input form allows turning the warp/flare on or off, while BGM1612 is hard-coded to be always on.
\end{enumerate}
\hspace{2mm} Overall, the Galaxia simulation results are tightly correlated with BGM1612 as expected. The bulge and disk kinematics are consistent with observed measurements from HST, and the addition of VLM stars in the bulge IMF is an advantage over Besan\c{c}on for surface density calculations at dimmer (and redder) magnitudes. However, the too low bar angle still persists in the current version.\\
\indent Finally, the N-body model that Galaxia implements is a self-consistent realization of the formation of the stellar halo in Milky Way-type galaxies via the formulation presented in \cite{bullock:2005aa}. The approach follows the evolution of accreted satellite galaxies in the halo formation process, and makes important distinctions between the evolution of light and dark matter within the host galaxies. The density of the accreted halo follows a varying power-law distribution, which changes radial slope from -1 within 10 kpc to -4 beyond 50 kpc. The distribution of stars is more centrally located in the halo compared to the dark matter distribution. This is expected for the stars building the stellar halo to be more tightly bound than the dark matter material that builds up the dark matter halo. The model is largely successful in reproducing the observed properties of surviving Milky Way satellites and the stellar halo. The characteristics of the inner bulge region are not incorporated in the N-body spawning of particles and is generally not suited for studying the most central regions of the Milky Way. For these reasons, we do not conduct an analysis of the N-body model with regard to the Stanek field.

\begin{figure*}
\figurenum{10}
\plotone{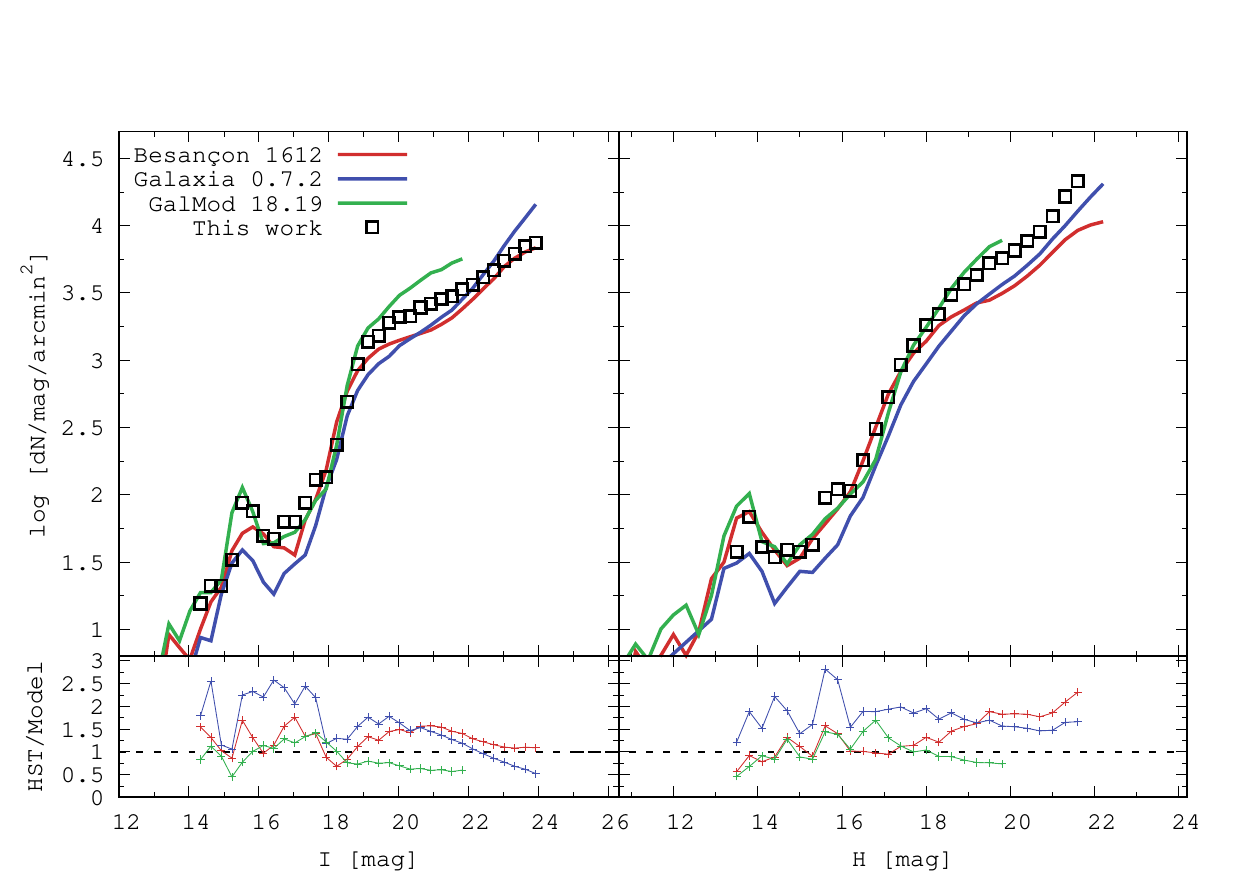}
\caption{\scriptsize{\textit{Top-left}: I-band LF of Stanek compared to the models. \textit{Lower-left}: Ratio of raw Stanek counts to the models raw counts for the corresponding surface densities. \textit{Top-right}, \textit{lower-right}: Similar comparison for the H-band data. The extinction is identical for each sample, however extinction-corrections are not applied here.}\label{fig:LF+Ratio}}
\end{figure*}

\begin{deluxetable}{cccccccc}
\deluxetablecaption{Observed Bulge Luminosity \label{table:5} Function\label{table:5}}
\tablecolumns{8}
\setlength{\tabcolsep}{2.5pt}
\tablenum{5}
\tablewidth{\columnwidth}
\tablehead{
\colhead{$V$} &
\colhead{log $N_{V}$} & \colhead{$I$} & \colhead{log $N_{I}$} & \colhead{$J$}& \colhead{log $N_{J}$} 
& \colhead{$H$}& \colhead{log $N_{H}$}
}
\startdata
14.629 & 0.484 & 14.329 & 1.192 & 14.1 & 1.284 & 13.5 & 1.575\\
14.929 & 1.262 & 14.629 & 1.323 & 14.4 & 1.363 & 13.8 & 1.837\\
15.229 & 1.086 & 14.929 & 1.322 & 14.7 & 1.626 & 14.1 & 1.612\\
15.529 & 1.505 & 15.229 & 1.517 & 15.0 & 1.359 & 14.4 & 1.537\\
15.829 & 1.630 & 15.529 & 1.941 & 15.3 & 1.474 & 14.7 & 1.592\\
16.129 & 1.907 & 15.829 & 1.878 & 15.6 & 1.481 & 15.0 & 1.576\\
16.429 & 1.982 & 16.129 & 1.694 & 15.9 & 1.688 & 15.3 & 1.628\\
16.729 & 1.774 & 16.429 & 1.673 & 16.2 & 1.930 & 15.6 & 1.979\\
17.029 & 1.614 & 16.729 & 1.798 & 16.5 & 2.003 & 15.9 & 2.041\\
17.329 & 1.820 & 17.029 & 1.799 & 16.8 & 2.132 & 16.2 & 2.029\\
17.629 & 1.679 & 17.329 & 1.940 & 17.1 & 2.394 & 16.5 & 2.256\\
17.929 & 2.025 & 17.629 & 2.111 & 17.4 & 2.564 & 16.8 & 2.491\\
18.229 & 2.097 & 17.929 & 2.128 & 17.7 & 2.629 & 17.1 & 2.724\\
18.529 & 2.255 & 18.229 & 2.373 & 18.0 & 2.856 & 17.4 & 2.966\\
19.129 & 2.495 & 18.529 & 2.691 & 18.3 & 3.072 & 17.7 & 3.110\\
19.429 & 2.849 & 18.829 & 2.968 & 18.6 & 3.195 & 18.0 & 3.261\\
19.729 & 3.042 & 19.129 & 3.138 & 18.9 & 3.294 & 18.3 & 3.340\\
20.029 & 3.093 & 19.429 & 3.181 & 19.2 & 3.394 & 18.6 & 3.484\\
20.329 & 3.185 & 19.729 & 3.279 & 19.5 & 3.484 & 18.9 & 3.564\\
20.629 & 3.239 & 20.029 & 3.321 & 19.8 & 3.564 & 19.2 & 3.633\\
20.929 & 3.247 & 20.329 & 3.326 & 20.1 & 3.618 & 19.5 & 3.721\\
21.229 & 3.254 & 20.629 & 3.392 & 20.4 & 3.695 & 19.8 & 3.758\\
21.529 & 3.260 & 20.929 & 3.420 & 20.7 & 3.747 & 20.1 & 3.816\\
21.829 & 3.280 & 21.229 & 3.455 & 21.0 & 3.817 & 20.4 & 3.886\\
22.129 & 3.283 & 21.529 & 3.476 & 21.3 & 3.888 & 20.7 & 3.954\\
22.429 & 3.287 & 21.829 & 3.528 & 21.6 & 3.976 & 21.0 & 4.071\\
22.729 & 3.286 & 22.129 & 3.562 & 21.9 & 4.049 & 21.3 & 4.216\\
23.029 & 3.290 & 22.429 & 3.616 & 22.2 & 4.159 & 21.6 & 4.328\\
23.329 & 3.305 & 22.729 & 3.669\\
23.629 & 3.342 & 23.029 & 3.740\\
23.929 & 3.357 & 23.329 & 3.792\\
24.229 & 3.388 & 23.629 & 3.846\\
24.529 & 3.402 & 23.929 & 3.873\\
\enddata
\tablenotetext{}{\footnotesize{\textbf{Notes}. Magnitude cutoff is at the 50\% completeness threshold. Units for star counts $N$ are consistent with that of Figure \ref{fig:LF+Ratio}, i.e. [d$N$/mag/arcmin$^{2}$].}}
\end{deluxetable}

\subsubsection{\normalfont{GalMod}} \label{sec:galmod}
GalMod\footnote[11]{\href{https://www.galmod.org/gal/}{\scriptsize https://www.galmod.org/gal/}} \citep{pasetto:2016baa, pasetto:2018aa, pasetto:2019aa} version \texttt{18.19} is the most recently published population synthesis model considered in this work. The simulation offers significantly more adjustable parameters over the previous models, which include the choice of 24 different photometric bands, fine-tuning of the density normalization factor, $\rho$, for each composite stellar population (CSP) (which also includes the ISM), 16 different combinations of SFR/IMF for each CSP, and the dark matter (DM) circular speed factor, scale radius, and flattening. Further, there are some distinct differences between GalMod and the previous models; the model incorporates a more realistic bulge-bar angle of $\sim$28$\degree$. GalMod generates convolved PDF's for star counts and then populates them with synthetic stars to obtain the CMD's and star counts, whereas Besan\c{c}on and Galaxia work in the opposite manner. GalMod's PDF-generating technique is particularly useful in the era of very large surveys that we are currently entering. Although this process differs from the prior models, quantitatively the GalMod approach to generating CMD's is identical to the other models. GalMod uses the geometry-independent ray-tracing extinction model of \cite{natale:2017aa} which is based on \cite{draine:2007aa}, and can realize a collisional or collisionless model generator for an N-body integrator.\\
\indent Additionally, there is a private version of the code that has tools to implement machine learning techniques for data-fitting convergence, uses dynamical estimators connected to the global Galactic potential, and other features. For an extensive comparison between GalMod and Besan\c{c}on, we refer the reader to \cite{pasetto:2016baa}. The remainder of this section will address the three discrepancies outlined in the previous sub-sections:
\begin{enumerate}
\item There are four different IMF's that can be used to simulate the Bulge/Bar CSP within GalMod; \cite{salpeter:1955aa}, \cite{scalo:1986aa}, \cite{kroupa:2001aa}, and \cite{chabrier:2003aa}. If no IMF is specified by the user, the model implements \cite{kroupa:2001aa} by default. This is a broken power-law with varying slopes; $\alpha = 2.3 \textrm{ for } 1.0M_{\sun} \leq m < \infty, \alpha = 2.7 \textrm{ for } 0.5M_{\sun} \leq m < 1.0M_{\sun}, \textrm{ and } \alpha = 1.8 \textrm{ for } 0.16M_{\sun} \leq m < 0.5M_{\sun}$. Figure \ref{fig:LF+Ratio} shows the GalMod predicted star counts compared to the Stanek field observations and other models. The overall shape of the LF agrees well in both I and H bands, while the model over-predicts star counts in the I-band range $I > 19$. The normalized H-band counts agree quite well with observations and are certainly the most consistent integrated across a majority of the magnitude range.
\item The GalMod kinematics exhibit a mean PM value consistent with prior results, however they show significant non-Gaussianity for the distributions in both $\mu_{l}$ and $\mu_{b}$ directions. The green data points in the bottom panel of Figure \ref{fig:pm-diag} show the PM distribution for the bulge RGB proxy stars from GalMod, which are clearly not well-fit by a Gaussian. Finally, the circular velocity, $V_{c}$, at the Sun location that GalMod uses is 220.8 km s$^{-1}$, which marginally smaller than the previous two models.
\item The $28\degree$ bulge-bar angle that Galmod uses is in better agreement with some prior observations. This angle will lead to proper descriptions of bulge-bulge and bulge-disk lensing probabilities, and a more realistic ratio of bulge-to-disk source stars for microlensing.
\end{enumerate}

\hspace{2mm} As stated above, the SFR in each CSP can take one of four different forms; constant, exponential, linear, or \cite{rosin:1933aa}. The \cite{rosin:1933aa} SFR is used by default and describes a rapid increase of the SFR up to a given time (free parameter), followed by a shallow decrease to the present day. This SFR has the functional form:

\begin{equation}
\psi (t) = \tau_{{(t_{1},t_{2})}} \psi_{0}t^{\beta}e^{-\frac{t}{h_{\tau}}},
\end{equation}

where $\beta$ is the power-law exponent ($1 \neq \beta > 0$) and $h_{\tau}$ is the timescale ($h_{\tau} > 1$). The prior models both use a single burst for their bulge SFR.\\
\indent Table \ref{table:5} shows the binned LF for each of the four observed wavelengths (V, I, J, H) and Figure \ref{fig:LF+Ratio} shows the Stanek field LF and the Galactic models in I-band and H-band. BGM1612 under-predicts star counts by $\sim$33\% at the RC location and integrated over the magnitude range 19.5-23.5 in I-band. Galaxia undercounts bulge stars more severely at brighter magnitudes. However, because of the accounting for VLM stars in the IMF, Galaxia counts more faint stars than BGM1612. This matches more closely to the observed Stanek field numbers and even over-counts at the faintest I-band end. In both plots, Galaxia overtakes BGM1612 in counts dimmer than the $0.15M_{\sun}$ low-mass cutoff (at $\sim$21 mag in I-band and $\sim$19 mag in H-band). The lower panels of Figure \ref{fig:LF+Ratio} show the raw surface density ratio between observation and the models. GalMod most closely predicts the star counts in both wavebands, with an accuracy of $\sim$10\% at I $<$ 20 and $\sim$5\% at H $<$ 19.\\
\indent With the larger bulge-bar angle, and higher predicted surface densities, GalMod stands to be a promising tool for further simulations and accurately predicting microlensing observables for WFIRST. The low-mass cutoff presents an issue for predicted surface densities at the dimmest magnitudes, as well as bulge star kinematics that are not in solid agreement with observations or other models. The purpose of this section (\ref{sec:model-compare}) is not to fully simulate the microlensing mission itself, but to conduct a detailed comparison of popular Galactic models to observations.
\section{Conclusion} \label{sec:6}
We have accurately measured PMs of the disk, bulge, and mixed stellar populations to within $\sim$20km s$^{-1}$ around $I=21$ and $H=20$ and within $\sim$60km s$^{-1}$ near the 50\% completeness limits of $I=24$ and $H=22$ and have likely probed the far-side disk population beyond the Galactic center. Our measured PM dispersions for each population are largely in agreement with prior studies, however, we measure a bulge component anisotropy ratio of $\sigma_{l}/\sigma_{b} = 1.03 \pm 0.04$ which is significantly smaller than prior results from studies in nearby bulge fields. We note that most of the past studies analyzed the `mixed' populations and are contaminated by disk rotation. The exception is the study of \cite{Calamida:2015aa}, who measure the pure-bulge anisotropy ratio in the SWEEPS field to be in agreement with our result. We applied magnitude-dependent scaling factors to the cleaned star counts in both I and H bands in order to properly account for bulge members that are excluded by our PM cut.\\
\indent The resulting bulge surface densities were compared to several Galactic population synthesis models. We find that the Besan\c{c}on and Galaxia models generally underpredict star counts in both wavebands from $33\%$ to upwards of $75\%$ in the most severe case (i.e $H > 18$). The bulge-bar angle that is smaller than previously measured values may be due to a degeneracy between this angle and the RC absolute magnitude dispersion, as discussed in \cite{simion:2017aa}.\\
\indent On the other hand, the newest population synthesis model, GalMod, produces bulge star counts that are in best agreement with the observed values along the Stanek sight-line. Although the model overpredicts bulge star counts by $\sim$35\% below the MSTO in I-band, it only overpredicts the counts by $\sim$5-10\% in the H-band data. These redder data are more closely aligned with the wavebands that the WFIRST microlensing mission will utilize during its observations.\\
\indent There are two drawbacks of note with GalMod; the bulge kinematics show a non-Gaussian PM distribution in both $\mu_l$ and $\mu_{b}$ directions. This disagrees with the measured kinematics of Stanek field observations presented in this paper as well as prior observational results in nearby bulge fields \citep{Clarkson:2008aa, Calamida:2015aa}. The second drawback is the low-mass cutoff of the empirical IMF; they do not allow the inclusion of the dimmest stars, which need to be factored in when attempting to simulate the lens and source star distribution and characterization in bulge sight-lines for WFIRST observations.\\ 
\indent The previous study of P19 worked to adjust the microlensing event rate based on a similar comparison to ours, between the SWEEPS field \citep{Calamida:2015aa} and an older version of the Besan\c{c}on model that used a shallow IMF and unrealistic bulge kinematics. The authors subsequently correct for these inaccuracies so that their simulations correctly predict microlensing event rates that match observations. Our star count results and comparison with the Galactic models support P19 with regard to kinematics and low-mass IMF corrections that are required. As P19 point out, there is an important need to advance simulation capabilities, particularly the correcting of stellar surface densities in highly-crowded bulge fields, in order to optimize WFIRST's observing strategy with regard to direct mass measurements. Lastly, an important precursor advancement we have also shown in this paper is an accurate description of the source magnitude distribution very nearby to the provisional WFIRST microlensing fields and to a (near-IR) depth overlapping what is achievable by WFIRST.

\subsection{\normalfont{Implications for WFIRST}} \label{sec:WFIRST Implications}
Over the four design reference mission studies, beginning with the mission proposed by the 2010 decadal survey, there has been significant variance in estimates of the expected microlensing event rate, optical depth and exoplanet yield. Calculations of these mission success criteria were based on disparate population synthesis models, measurements along sight-lines distant from the proposed WFIRST microlensing fields, and shallow LF's that do not overlap the WFIRST wavebands. In this work, we have measured the stellar populations directly adjacent to the WFIRST microlensing fields to near-IR wavebands J and H. These measurements take the form of kinematic distributions and dispersions in the field, as well as accurately measured stellar surface densities which are compared to several population synthesis models.\\
\indent These results can be used to directly answer mission-critical scientific needs stated in \cite{spergel:2013aa} and \cite{Yee:2014aa}. Particularly, Yee et al. Section 1.2: \textit{Improve characterization of the WFIRST fields}, as well as Spergel et al., the \textit{scientific need to measure the: source star luminosity function, near-IR event rate, and relative bulge-to-disk planet frequency}. With regard to the Galactic models used in this work, We re-iterate P19 section 6.4: there is still room for improvement to Galactic models, we can put tighter constraints on microlensing observables and source/lens properties with updated models. Particularly, new PM results from Gaia DR2 \citep{brown2018gaia} can be included in the models to better characterize the kinematics within the WFIRST microlensing fields. There are plans to include these new PM measurements in a future GalMod release (S. Pasetto private comm). Additionally, the future Gaia DR3 and DR4 releases offer prospects of high accuracy positions and kinematics for the farthest stars yet toward the Galactic center at $\sim$8 kpc. The stellar properties within the models as well as future WFIRST preparatory work will benefit greatly from these future releases.\\
\indent Our PM analysis presented here shows that the stars along the sightline to Stanek's Window in the bulge and disk exhibit a PM (longitude and latitude, combined) of about 4.2 mas/yr.  While WFIRST will observe the Galactic bulge for six seasons over five years during its primary mission, it will not visit the same field with the same orientation for about 3.5 years, corresponding to $\sim$15 mas total PM per detected event.  This is the case, however, only for bulge-disk lensing. When compared to WFIRST's pixel scale of 110 mas, we may expect a typical star to exhibit 0.14pix of motion from one visit to the next. Using the color-dependent centroid shift method \citep{bennett:2006aa, bennett:2015aa, bhattacharya:2018aa} as well as the image elongation method \citep{bennett:2007aa}, we may expect a precision of 11\% on the lens-source separation with 3.5 years of baseline.\\
\indent Finally, the near-IR source magnitude distribution and other results presented here can be be combined with future studies to further simulate the scientific yields of the WFIRST microlensing survey. For instance, the previous extrapolation errors in the GULLS simulation software \citep{Penny:2013aa} can now be mitigated with newer, corrected models. \\
\\
\indent The authors would like to thank Stefano Pasetto for GalMod technical assistance. We also thank Annalisa Calamida, Kailash Sahu, and No\'e Kains for practical help and assistance. Finally, we thank the anonymous referee for constructive comments that led to a stronger manuscript.\\
\indent This work was performed in part under contract with the Center for Research and Exploration in Space Sciences \& Technologies (CRESST-II), which is a collaboration between NASA GSFC and surrounding universities. This work makes use of data products from GO-proposals 11664 and 12666 of the NASA/ESA Hubble Space Telescope, obtained by STScI. STScI is operated by the Association of Universities for Research in Astronomy, Inc., under NASA contract NAS5-26555. DPB and AB were supported by NASA through grant NASA-80NSSC18K0274.\\
\indent This publication makes use of the software packages: Astropy \citep{astropy:2018aa}, gnuplot, KS2 \citep{anderson:2008aa}, Matplotlib \citep{hunter:2007aa}, and Numpy \citep{oliphant:2006aa}.

\newpage
\bibliographystyle{aasjournal}
\bibliography{terry_2019_v0.6.bib}

\end{document}